\documentclass[fleqn,usenatbib]{rasti}

\usepackage{newtxtext,newtxmath}
\usepackage[T1]{fontenc}

\DeclareRobustCommand{\VAN}[3]{#2}
\let\VANthebibliography\thebibliography
\def\thebibliography{\DeclareRobustCommand{\VAN}[3]{##3}\VANthebibliography}

\usepackage{graphicx}	% Including figure files
\usepackage{amsmath}	% Advanced maths commands
\usepackage{ulem}
\usepackage[dvipsnames]{xcolor}

\newcommand{\orcid}[1]{\text{\href{https://orcid.org/#1}{\includegraphics[width=8pt]{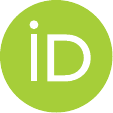}}}}

\newcommand{\cross}[1][1pt]{\ooalign{%
  \rule[1ex]{1ex}{#1}\cr% Horizontal bar
  \hss\rule{#1}{.7em}\hss\cr}}% Vertical bar

\title[Design and performance of a portable CBP]{Design and performance of a Collimated Beam Projector for telescope transmission measurement using a broadband light source}

\author[K. Sommer et al.]{
K. Sommer$^\orcid{0009-0007-5504-5838}$,$^{1}$\thanks{E-mail: kelian.sommer@umontpellier.fr}
J. Cohen-Tanugi$^\orcid{0000-0001-9022-4232}$,$^{1,2}$
B. Plez$^\orcid{0000-0002-0398-4434}$,$^{1}$
M. Betoule$^\orcid{0000-0003-0804-836X}$,$^{3}$
S. Bongard$^\orcid{0000-0002-3399-4588}$,$^{3}$
L. Le Guillou$^\orcid{0000-0001-7178-8868}$,$^{3}$
\newauthor
J. Neveu$^\orcid{0000-0002-6966-5946}$,$^{3,4}$
E. Nuss,$^{\cross, 1}$
E. Sepulveda,$^{3}$
T. Souverin,$^{3}$
M. Moniez$^\orcid{0000-0001-8716-6561}$,$^{4}$
and C. W. Stubbs$^\orcid{0000-0003-0347-1724}$,$^{5,6}$
\\
$^{1}$Laboratoire Univers et Particules de Montpellier, Université de Montpellier, CNRS, Montpellier, France\\
$^{2}$Laboratoire de Physique de Clermont, Université Clermont Auvergne, CNRS, F-63000 Clermont-Ferrand, France\\
$^{3}$LPNHE, CNRS/IN2P3 \& Sorbonne Université, 4 place Jussieu, 75005 Paris, France\\
$^{4}$Université Paris-Saclay, CNRS, IJCLab, 91405 Orsay, France\\
$^{5}$Department of Physics, Harvard University, 17 Oxford street, Cambridge, MA 02143, USA\\
$^{6}$Department of Astronomy, Harvard University, and Center for Astrophysics, 60 Garden Street, Cambridge, MA 02143, USA
}

\date{Accepted XXX. Received YYY; in original form ZZZ}

\pubyear{2023}

\begin{document}
\label{firstpage}
\pagerange{\pageref{firstpage}--\pageref{lastpage}}
\maketitle

\begin{abstract}
% This is a simple template for authors to write new RASTI papers.
% The abstract should briefly describe the aims, methods, and main results of the paper.
% It should be a single paragraph not more than 250 words.
% No references should appear in the abstract.
Type Ia supernovae are the most direct cosmological probe to study dark energy in the recent Universe, for which the photometric calibration of astronomical instruments remains one major source of systematic uncertainties. To address this, recent advancements introduce Collimated Beam Projectors (CBP), aiming to enhance calibration by precisely measuring a telescope's throughput as a function of wavelength.
This work describes the performance of a prototype portable CBP. The experimental setup consists of a broadband Xenon light source replacing a more customary but much more demanding high-power laser source, coupled with a monochromator emitting light inside an integrating sphere monitored with a photodiode and a spectrograph. Light is injected at the focus of the CBP telescope projecting a collimated beam onto a solar cell whose quantum efficiency has been obtained by comparison with a NIST-calibrated photodiode. The throughput and signal-to-noise ratio achieved by comparing the photocurrent signal in the CBP photodiode to the one in the solar cell are computed.
We prove that the prototype, in its current state of development, is capable of achieving 1.2 per cent and 2.3 per cent precision on the integrated g and r bands of the ZTF photometric filter system respectively, in a reasonable amount of integration time. Central wavelength determination accuracy is kept below $\sim$ {0.91} nm and $\sim$ {0.58} nm for g and r bands. The expected photometric uncertainty caused by filter throughput measurement is approximately 5 mmag on the zero-point magnitude.
Several straightforward improvement paths are discussed to upgrade the current setup.
\end{abstract}

% Include between one and six keywords.

%%%%%%%%%%%%%%%%%%%%%%%%%%%%%%%%%%%%%%%%%%%%%%%%%%

%%%%%%%%%%%%%%%%% BODY OF PAPER %%%%%%%%%%%%%%%%%%

\section{Introduction}

% This is a simple template for authors to write new RASTI papers.
% See \texttt{rasti\_guide.tex}
% for a full user guide.

% All papers should start with an Introduction section, which sets the work
% in context, cites relevant earlier studies in the field by \citet{Fournier1901},
% and describes the problem the authors aim to solve \citep[e.g.][]{vanDijk1902}.
% Multiple citations can be joined in a simple way like \citet{deLaguarde1903, delaGuarde1904}.

Type 1a supernovae (SNe Ia) are a class of supernovae that are thought to result from the explosion of a white dwarf (WD) star in a binary system (see \citet{Hillebrandt2000} and \citet{Mazzali2007} for detailed reviews of SNe Ia physics). Their uniform intrinsic brightness at the 15 per cent level after standardization and their luminosity around $10^{10} L_{\odot}$ make them particularly useful for cosmology, in particular, to accurately measure the expansion rate $H(t)$ and to study the properties of dark energy with the determination of the equation of state parameter $w$ \citep{Garnavich1998, Efstathiou1999, Copeland2006}.
As the number of surveys increases, the statistical uncertainty on cosmological parameters measured with SNe Ia diminishes \citep{Brout2022}. 
For instance, the Vera Rubin Observatory will observe over 10,000 SNe Ia in its 10-year survey \citep{lsstsciencebook}. Systematic uncertainties thus become predominant, as demonstrated by \citet{Scolnic2018}. These include SNe Ia models, bias selections, galactic extinction, peculiar velocities, local environment, and not the least calibration errors. Indeed, instrumental photometric flux calibration represents the main source of uncertainty for the estimated distances of SNe Ia \citep{Stubbs2015} and for cosmological parameters inferred with this probe. We refer the reader to \citet{Goobar2011} for a review of SNe Ia cosmology and to \citet{Betoule2014, Scolnic2014} for a review of calibration uncertainties.

This flux calibration issue can be split in two different components: (i) the terrestrial atmospheric transmission, and  (ii) the wavelength-dependent instrumental {throughput}. The first component can be determined by comparing the measured flux of reference stars in the field of view for each photometric band with catalogs of spectrophotometric reference stars. These ultimately rely on 3 pure hydrogen white dwarfs primary CALSPEC standards \citep{Bohlin2020} for which physical models describe very accurately the spectra \citep{Narayan2019}. Models are internally consistent with Hubble Space Telescope photometry to better than 1 per cent (see \citet{Bohlin2014_2} for an extended review of absolute calibration methods). 
To isolate the uncertainty component due to the instrument throughput, multiple efforts with in situ artificial methods have emerged for different surveys \citep{Stubbs2006, Stubbs2010, Doi2010, Marshall2013, Regnault2015, Lombardo2017}. They involve laboratory metrology standards with extremely well-characterized photodetectors.

\citet{Stubbs2006} proposed a method utilizing tunable lasers to characterize the relative wavelength response of a complete imaging setup relying on a flat-field screen and a calibrated reference photodiode. It was later implemented to determine the relative throughput of the PanSTARRS telescope and the Gigapixel imager with total uncertainty higher than the per cent level, all systematics considered \citep{Stubbs2010}. \citet{Doi2010} developed a monochromatic illumination system ~--~ which consisted of a broadband lamp, a monochromator, an integrating sphere and a photodiode ~--~ dedicated towards in situ measurements of the response function of the mosaicked CCD imager used in the Sloan Digital Sky Survey (SDSS). They were successfully able to quantify response variations to be lower than 0.01 mag for SDSS $griz$ bands over the entire duration of the survey. Their conclusion stated that the main driver behind throughput evolution with time was the seasonal variations described by temperature effect. Therefore, as it is planned for the Vera Rubin Observatory, the throughput measurement operation should be performed regularly across the year. \citet{Marshall2013} developed the DECal instrument which provided both broadband flat fields and narrowband spectrophotometric calibration for the Dark Energy Camera (DECam, see \citealt{DECam2015}) installed on the 4-m Blanco Telescope of the Dark Energy Survey (DES). The flat field system employed LEDs to illuminate individual photometric filters, while the spectrophotometric calibration system involved a standard tunable light source from a monochromator that was directed onto the flat field screen using a specialized fiber bundle and diffuser. Calibrated photodiodes placed along the beam track the telescope's throughput across wavelengths. They successfully deployed a prototype at the Swope 1-m and du Pont 2.5-m telescopes at Las Campanas Observatory in Chile and measured the throughput of the $ugriBVYJHK$ filters used in the WIRC and RetroCam instruments during the Carnegie Supernova Project with an accuracy of 1\%. The DICE (Direct Illumination Calibration Experiment, see \citealt{Regnault2015}) instrument was composed of multiple narrow-spectrum LEDs ranging from UV to NIR spectral range emulating a point-like source placed at a finite distance from the telescope entrance pupil, yielding a flat field illumination that covered the entire field of view of the imager. They were able to measure the MegaCam and SkyMapper $uvgriz$ passbands normalisations terms relative to $r$ band with lower than 0.5\% relative uncertainty. \citet{Lombardo2017} built the SNIFS Calibration Apparatus (SCALA) system for the SuperNova integral field spectrograph (SNIFS) which used a monochromator-based instrument with integrated spheres whose output exit ports were collimated and projects large spots onto the focal plane array of the SNIFS instrument. They assessed the long-term repeatability of the light emitted by SCALA to be better than 1\% and evaluated the color-calibration stability of the University of Hawaii 2.2-m (UH 88) telescope, which remained at the per cent level over one year across the wavelength range from 400 nm to 900 nm.

We focus here on the development of a calibration device, called a Collimated Beam Projector (CBP), designed to accurately measure the wavelength-dependent response of the instrument $R_{tel}(\lambda)$, including the optics, the CCD camera, and the filters.
A first design was proposed by \citet{Coughlin2016, Coughlin2018}. Their prototype was tested on the CTIO 0.9-m telescope and the DECam wide field imager.
The principle and main advantage of the CBP compared to previously cited instruments, is that it relies on illuminating one or multiple spots of the primary mirror with a stable collimated monochromatic light beam. The incident wavefront is parallel upon entering the telescope, resulting in the creation of a collimated beam that covers a fraction of the entire pupil. The full-pupil illumination is synthesized by scanning at multiple angles and radii. Consequently, this design minimizes reflections and stray light within the optical setup, which is a considerable improvement over utilizing a flat-field screen for illumination \citep{Coughlin2016}. It also facilitates the separation of direct illumination beams and ghost spots for analysis and computations. Finally, it measures the output on the detector, which allows the calculation of the throughput. The total transmission function is then rasterized by interpolating between angles and spot positions. 

\begin{figure}
    \centering
    \includegraphics[width=\columnwidth]{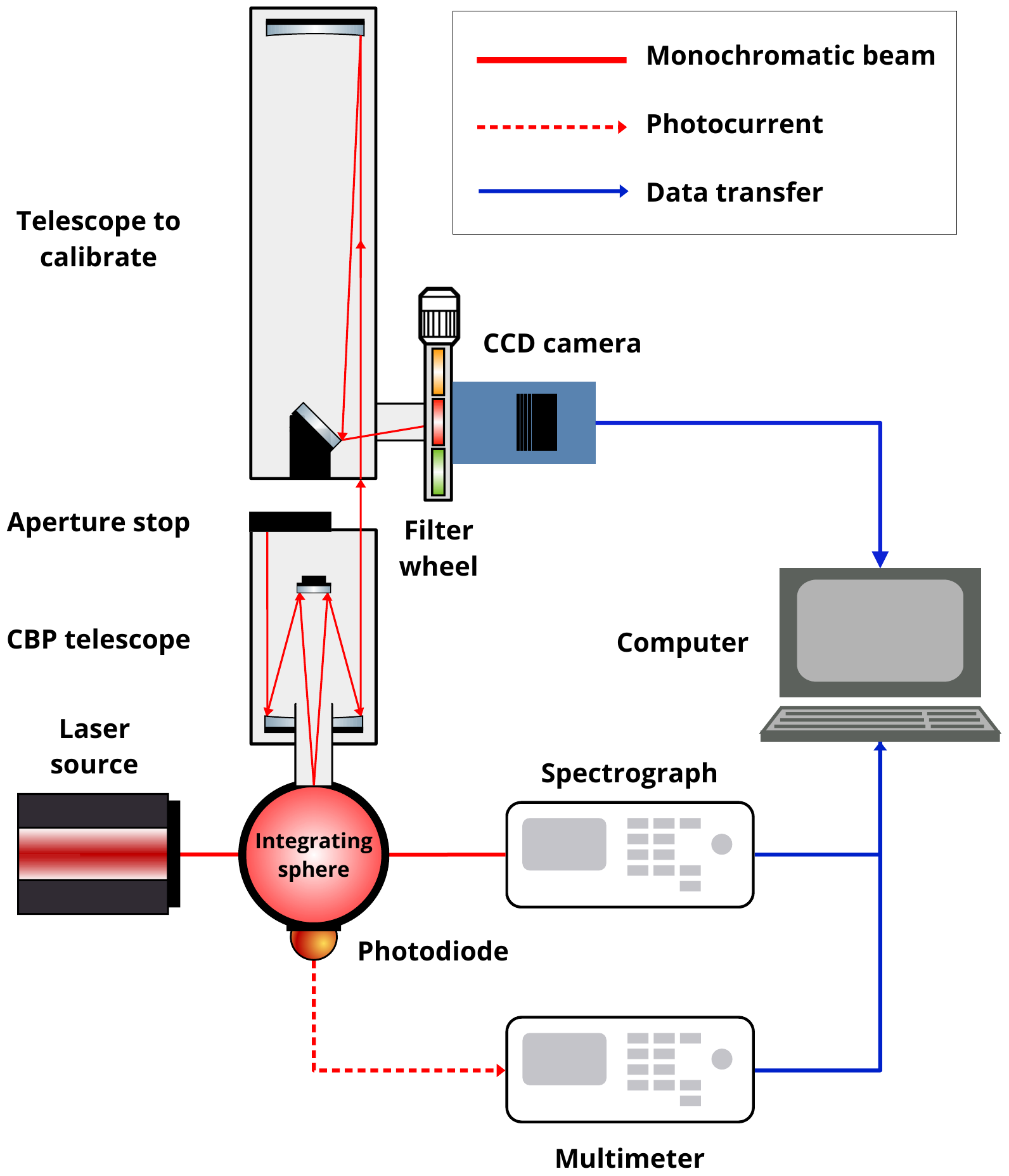}
    \caption{Sketch of the CBP used by \citet{Souverin2022} to calibrate the StarDICE 400 mm F/3.6 Newtonian telescope.  The CBP telescope acts as a projector to illuminate the {StarDICE} telescope with a collimated beam.
    The laser generates a monochromatic beam in the range 
    300-1100\,nm, monitored by a calibrated spectrograph. The flux level scattered inside the integrating sphere is measured by the photodiode. 
    A fraction of the light {in the integrating sphere} is injected into the telescope to be calibrated through the CBP telescope, whose throughput was measured beforehand with a solar cell. 
    An aperture-stop is added {to prevent the beam from encountering} the StarDICE secondary mirror holder. The sensor, here a CCD camera, is {placed at the focal plane} to determine the actual response in terms of $\mathrm{ADU\,W\,m ^{-2}\,nm^{-1}}$ in an observation-like position through all photometric filters. {This experiment aims to measure the StarDICE telescope throughput in $\mathrm{ADU\,photon^{-1}}$, since the number of photons injected into the StarDICE telescope by the CBP is known.}}
    \label{fig:original_cbp_layout}
\end{figure}

Recently, \citet{Mondrik2023} designed another CBP prototype to measure the StarDICE \citep{Betoule2023} proof-of-concept 254-mm telescope, achieving a $\sim$ 3 per cent accuracy over the 300 - 1000\,nm wavelength range. 
\citet{Souverin2022} built an enhanced version of the CBP for the final StarDICE experiment (Fig.~\ref{fig:original_cbp_layout}) and were able to measure the response of a 400 mm F/4 Newtonian telescope coupled with {\textit{ugrizy}} filters and CCD sensor at the 0.1 per cent statistical uncertainty level.
This version required a powerful \textit{class 4} laser source, unpractical for transportation and exploitation due to its fragility and security concerns. The same kind of design will be used for the Vera C. Rubin Observatory CBP currently in construction, with a laser packed inside a dedicated thermally controlled enclosure \citep{Ingraham2022}.
For the Dark Energy Science Collaboration (DESC) 10-year supernovae experiment \citep{DESC} of the Vera C. Rubin Observatory, systematic uncertainties on the $\textit{griz}$ bands zero points and central wavelengths are required to be lower than 1 mmag and 0.1 nm respectively. This is undertaken with the aim of harnessing the capabilities of the instrumentation and survey, leading to substantial enhancements in cosmological constraints. These metrics evaluated on the Zwicky Transient Facility (ZTF) photometric system bands ztf-\textit{g} and ztf-\textit{r} \citep{Bellm2019, Dekany2020} will serve as benchmarks throughout the remainder of this study. Given that ZTF neither possessed nor has plans to acquire a permanent CBP for bandpasses throughput measurements, there exists a unique opportunity to design an innovative solution that fulfills an actual requirement.

Therefore, this work focuses on the first attempt to develop a ``traveling CBP'', a transportable version of a CBP for on-site measurements at multiple observatories. We substitute the sensitive tunable laser with a more conventional light source. The primary motivation for switching to broadband sources is their simplified maintenance, cost-effectiveness, and reduced susceptibility to environmental flaws compared to lasers. This versatility renders them more adaptable than lasers, the challenge now being to reach a sufficiently high monochromatic brightness.

\begin{figure*}
    \centering
    \includegraphics[width=1.0\linewidth]{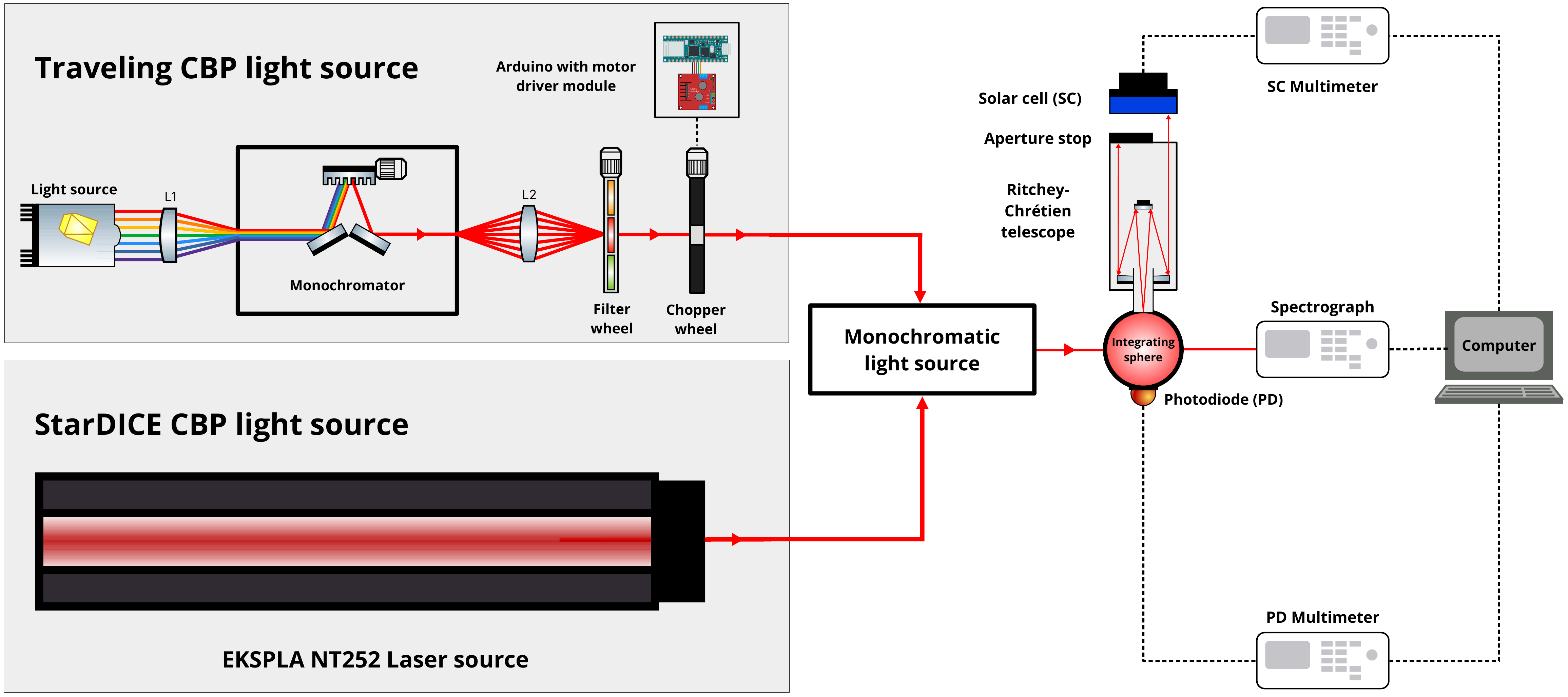}
    \caption{Schematics of the experimental setup to measure the CBP telescope throughput. {On the left, either a class 4 laser source (below) or the more standard light source coupled with a monochromator (studied in this work) delivers a monochromatic light beam.} This beam is injected into the integrating sphere (IS) in a free-space configuration, with a photodiode (PD) monitoring the light flux. The light escapes from the IS through a hole placed at the focus of the CBP telescope operating in reverse mode. After reflections on the secondary and primary mirrors, it emerges as a parallel beam through a mask before reaching the solar cell (SC). Two precision electrometers are used to measure the charge accumulation from the SC and the PD. Elements are not to scale.}
   \label{fig:schematics_equipment_final} 
\end{figure*}

The paper is structured as follows. In Section \ref{sec:experimentalsetup}, we describe the architecture of the setup and explain the role of each component. In Section \ref{sec:data}, we detail the data acquisition procedure, and the noise model and provide a complete analysis with forward modelling techniques. Section \ref{sec:results} presents the CBP throughput curve and other relevant metrics. In Section \ref{sec:measurement_uncertainty}, we address the budget measurement uncertainties including systematics.
In Section \ref{sec:discussion}, we discuss the results, potential future improvements, and methods to mitigate some measurement and systematic uncertainties. 
We draw our conclusions in Section \ref{sec:conclusions}.

\section{Experimental setup}
\label{sec:experimentalsetup}

\begin{figure*}
  \centering
  \includegraphics[width=1.0\linewidth]{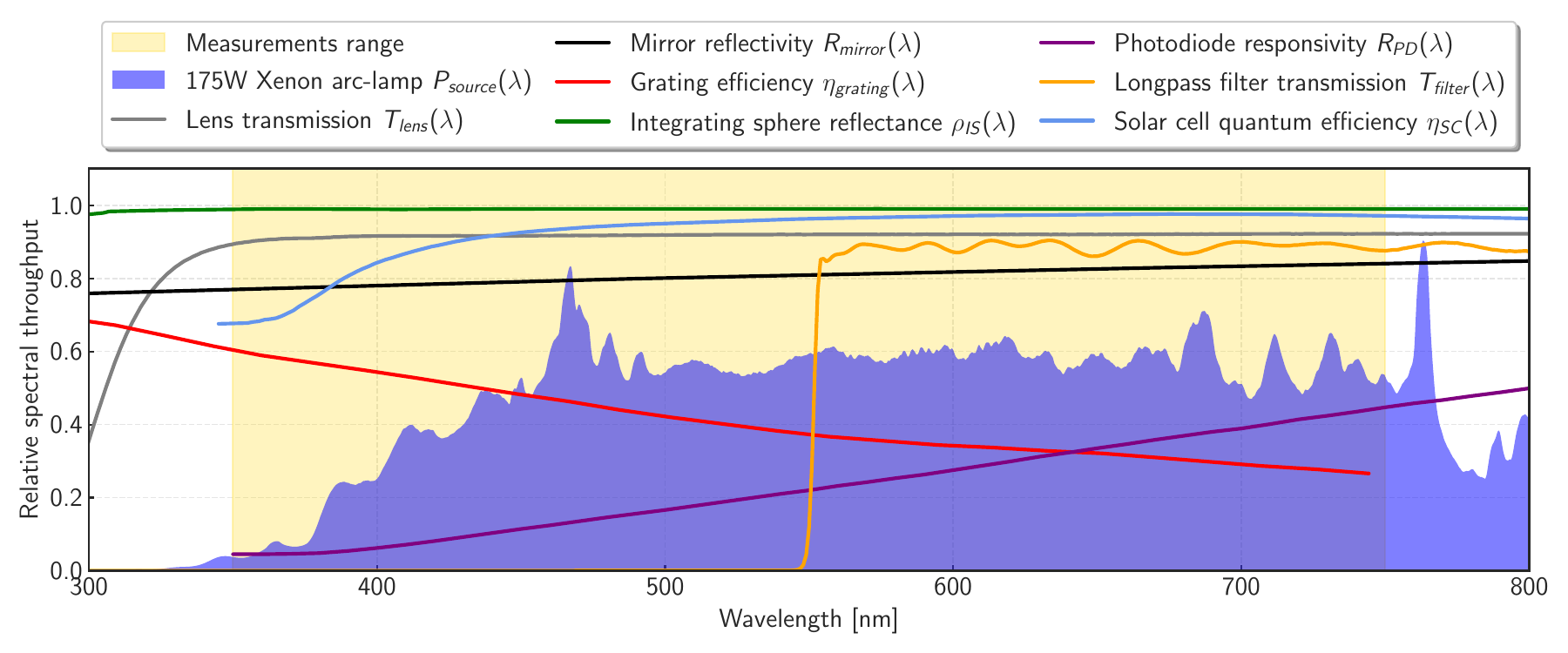}
  \caption{Relative spectral throughput of each element constituting the instrument. The golden area shows the wavelength range of this study. With the exception of the quantum efficiency of the SC (light blue) measured in $e^{-}/\gamma$ relative to the NIST-calibrated photodiode \citep{Brownsberger2022}, all the data is obtained from the specifications provided by manufacturers.}
  \label{fig:spectrum_throughputs}
\end{figure*}
The experimental setup used in this work is shown in Fig.~\ref{fig:schematics_equipment_final}, where the elements of a CBP (Fig.~\ref{fig:original_cbp_layout}) can be recognized.
In order to calibrate the CBP telescope and its attached photodiode (PD), we illuminate a solar cell (SC) at the exit of the CBP telescope. This SC has been calibrated using NIST standard calibrators \citep{Larason2008, Brownsberger2022} and {has served for the StarDICE telescope calibration}.
Our goal is to calibrate the throughput of the CBP telescope, using another source than the laser, and relate it to the PD current. 
{The main challenge stems from the fact that the solar cell requires a high illumination intensity, easily reachable with a class 4 laser, but not necessarily with a more standard light source}.

In this section, we provide technical details of each element in the instrumental setup. 

\subsection{Traveling CBP light source prototype}\label{sec:light_source}

We employ a standard broadband 175W Xenon arc lamp (Spectral Products ASB-XE-175) of 5600K color temperature, built with a Xenon parabolic lamp. Its emission spectrum, which lies in the range 200-2000 nm, is shown in Fig.~\ref{fig:spectrum_throughputs}. This broadband light is then focused into a Czerny-Turner (Cornerstone 130 1/8\,m) monochromator entrance slit with an uncoated lens located at a known distance of the entrance slit. The focal length of the monochromator is 125\,mm for an acceptance aperture of F/3.7. The grating available in the monochromator used in this test is a 30x30\,mm$^2$ with 2400\,grooves/mm and a blaze wavelength of 275\,nm. The manufacturer's documentation recommends setting the entrance and exit slits at approximately 1\,mm width for a 2-3\,nm spectral bandwidth. Slit heights are adjusted to the maximum of 12\,mm to collect as much {light} as possible. Multiple diffraction orders can coexist in the beam at the exit slit. Therefore, {a long pass filter designed to block undesirable higher diffraction orders} is inserted between the exit slit of the monochromator and the chopper wheel {(introduced below)}. The transmission curve of this Thorlabs 550\,nm Hard-Coated Edgepass Filter, as provided by the manufacturer, is shown in Fig.~\ref{fig:spectrum_throughputs}.

\begin{figure}
    \centering
        \includegraphics[width=\columnwidth]{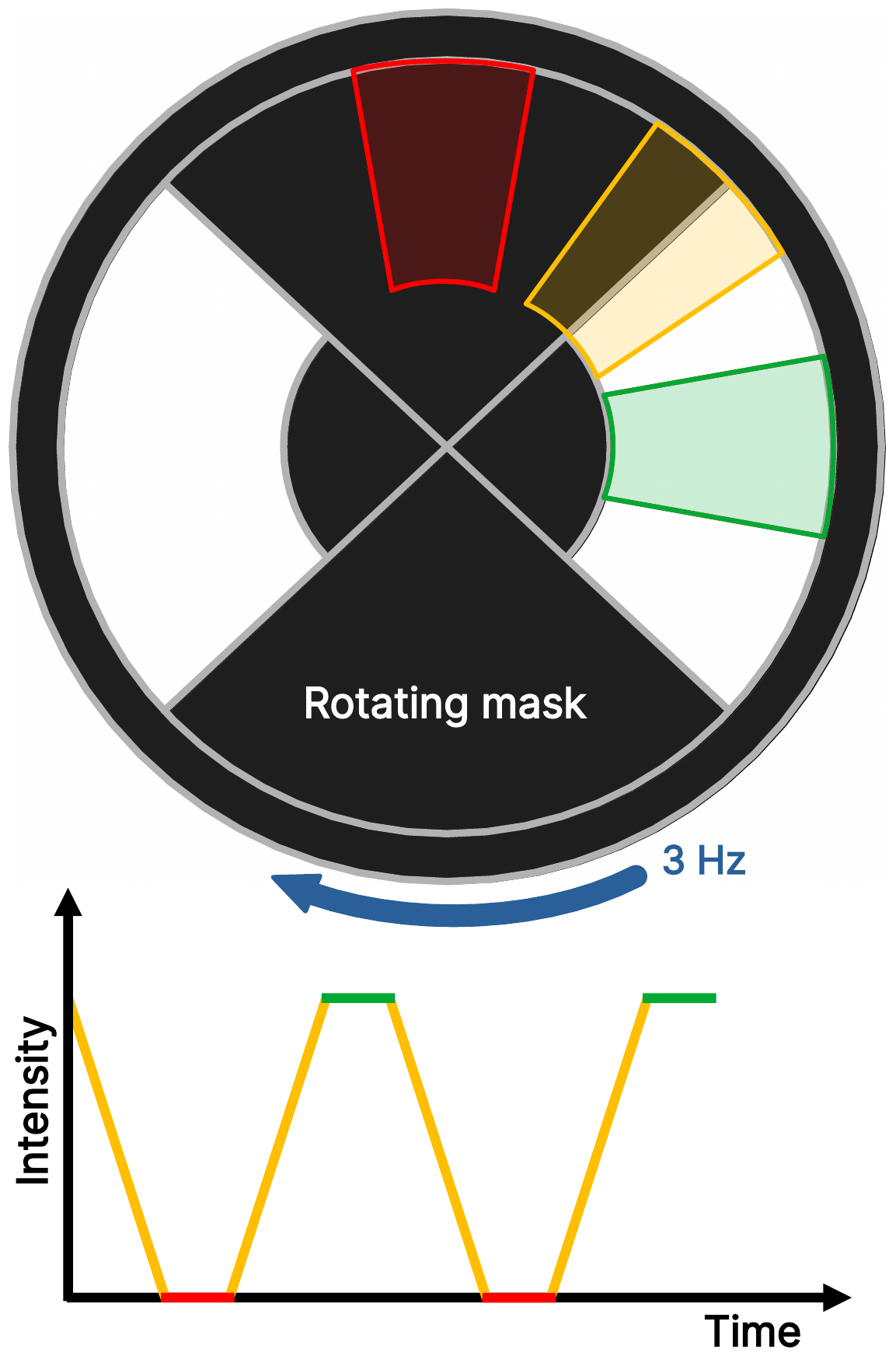}
    \caption{\textit{Top panel}: Sketch of the chopper wheel. The mask has 2 blades (black), rotates at 3 Hz, and lets the beam travel through the aperture (colors). Green: fully open aperture state, orange: partially open aperture state; red: fully closed aperture state. \textit{Bottom panel}: sketch of a time series resulting from the rotation of the chopper wheel; the colors correspond to the situations depicted in the top figure.
    }
    \label{fig:chopper_mask}
\end{figure}

Next, a chopper wheel, made of a simple stepper motor driven by an Arduino Nano micro-controller coupled with a ULN2003A unipolar motor driver, rotates a custom 3D printed mask with two apertures in front of a fixed slit (see Fig.~\ref{fig:chopper_mask}). Its purpose is to significantly decrease the level of $1/f$ noise that originates from small fluctuations of the operational amplifier bias voltage which can draw current through the relatively low shunt resistance of the solar cell. The amplitude of the contamination is inversely proportional to the resistance of the solar cell used in this study. The motor rotates at a rate of $\sim3$ Hz. The full analysis that consists of reducing the 1/f noise contribution, accounting for small rotation speed variations and aperture differences is presented in Section~\ref{sec:data}.

\subsection{Integrating sphere and spectrograph}

We use the General-Purpose Ø50 mm Integrating Sphere (IS) from Thorlabs in order to ensure identical beam fluxes toward the sensors. Its typical reflectance is better than 98 per cent in the range 350-1500 nm (see Fig.~\ref{fig:spectrum_throughputs}). A spectrograph connected to the IS through an optic fiber cable  {monitors spectral shapes of the monochromatic beams exiting the monochromator slit}. 

\subsection{Photodiode}
We use the SM05PD1B silicon-based photodiode (PD) from Thorlabs, designed to operate in the UV-VIS-NIR range (350-1100 nm), in order to measure the incoming light source intensity. 
It has a photosensitive area of 3.6 $\times$ 3.6 mm$^{2}$. The PD operates in photovoltaic mode to reduce dark current which is important to optimize performance with this type of low-illumination experiments. Electrical charges are sampled by a Keithley~6514 at 50\,Hz rate, in conjunction with a 1 power line cycle to get rid of the 50Hz power line influence.

During an integration {period}, the PD achieves charge levels ranging from 10 to 50 nC. The electrometer's bias current is less than 4 fA, which is negligible in comparison to average photocurrent levels of $\sim$ 20\,pA. A 1\,mm hole is placed in front of the photodiode in order to ensure that measurements are within the limits of the electrometer measurement range and do not saturate.

\subsection{CBP telescope and Solar Cell}
The CBP telescope was chosen to be compact and with the simplest optical layout in order to avoid internal reflections with field correctors or chromatic aberrations with lenses, which lead to the appearance of \textit{ghost patterns} (see \citet{Murray1949} for an in-depth explanation). It is an Omegon Ritchey-Chrétien equipped with two hyperbolic BK7 mirrors averaging 92.7 per cent reflectivity. It has a 152\,mm aperture and a 1370\,mm focal length. The primary mirror is larger than the SC. {A mask is mounted at the entrance of the telescope in order to only illuminate the SC photosensitive surface. Multiple tinfoil layers are added to ensure that the assembly is fully light-tight in the optical and infrared range.}

The solar cell is a single high-efficiency Sunpower C60 version already studied in \citet{Brownsberger2020, Brownsberger2022}. This SC offers a high quantum efficiency (QE) $>90per cent$ and a large collecting area $>10 \times 10$\,cm$^2$ at a low cost with great portability. Its output charge is measured by a Keysight~2987A multimeter at a sample rate of 500\,Hz, based on the acceptable SC time response (at least 200\,Hz). \citet{Brownsberger2022} review the absolute calibration procedure relative to the NIST at 0.1 per cent accuracy of the QE between 350 and 1100\,nm. One important downside of the SC is the strong dark signal caused by a low shunt resistance $R_{sh}${, which is due to manufacturing defects and causes power losses in solar cells by providing an alternate current path for the photocurrent}. Our model of SC was measured with $R_{sh} = 1.8$\,k$\Omega$, to be compared with the serial resistance $R_{ser} = 170$\,m$\Omega$, which is the electrical resistance that hinders the flow of current through the main current path in a solar cell. It is the same SC sample as the one used in \citet{Souverin2022}.

\section{Data acquisition and analysis}
\label{sec:data}
The monochromator was used to scan wavelengths from 350\,nm to 750\,nm with a 2\,nm step. With the spectrograph, spectra were obtained at five different wavelengths across the considered range of measurements to assess the quality of the monochromatic light beam (shown in Appendix \ref{sec:appendix_spectra}). We correct the wavelength values read on the monochromator using a third order polynomial fit to the 5 values from the spectrograph. 

At each wavelength, the charge accumulation from the SC and PD was read during a 20\,s integration time, resulting in $10^4$ and $10^3$ values, respectively. The monochromator shutter was closed, and the electrometer charges reset between each integration of 20\,s. Intertwined to this "signal" sequence, 4 sequences of dark current exposure measurements were performed by leaving the monochromator's shutter closed during the 20\,s integration time. We compute the average of the dark current for the PD and subtract it from each time series as it is estimated to behave like stationary noise. The sequence of charges was finally turned into time series of currents by finite difference, using the timestamps provided by the electrometer's clock. As an illustration, Fig.~\ref{fig:pd_raw_signal} shows in red the 20\,s signal in the PD for the $\lambda=550.5$\,nm setting. The 6\,Hz periodicity induced by the chopper is visible.

For the analysis, we adopted a purely modeling-based approach, favoring it over a lock-in detection method for several reasons. Firstly, the prototype bench had a limited existence of less than three weeks in Summer 2022, constrained by material and time limitations. Secondly, the lock-in detection method demands more effort for signals that deviate from being squared or sinusoidal. Lastly, its implementation becomes challenging when dealing with sensors and electrometers with varying bandwidths. The decision to configure the electrometers in charge mode was influenced by replicating methods from previous CBP versions \citep{Souverin2022, Mondrik2023}. This choice aimed to better account for the laser's pulse-to-pulse instability, often reaching 10\% or more. The chosen hardware integration allowed them to avoid worrying about the fact that time sampling (50 or 500 Hz) was well below the photodiode response time (\textmu s) to laser pulses (ns). Despite experimenting in current mode on the Keysight, where the SC is connected, contributing significantly to the uncertainty budget, no observable improvement was noted.

\subsection{Photodiode noise model}
The PD is affected by readout noise that we estimate experimentally by taking the standard deviation of 4 concatenated dark exposures: $\sigma_{readout} = 0.494$\,pA. It amounts at worst to 5 per cent of the PD signal in our dataset. This encompasses the combination of both dark noise and electrometer read-noise.

\begin{figure*}
  \centering
  \includegraphics[width=1.0\linewidth]{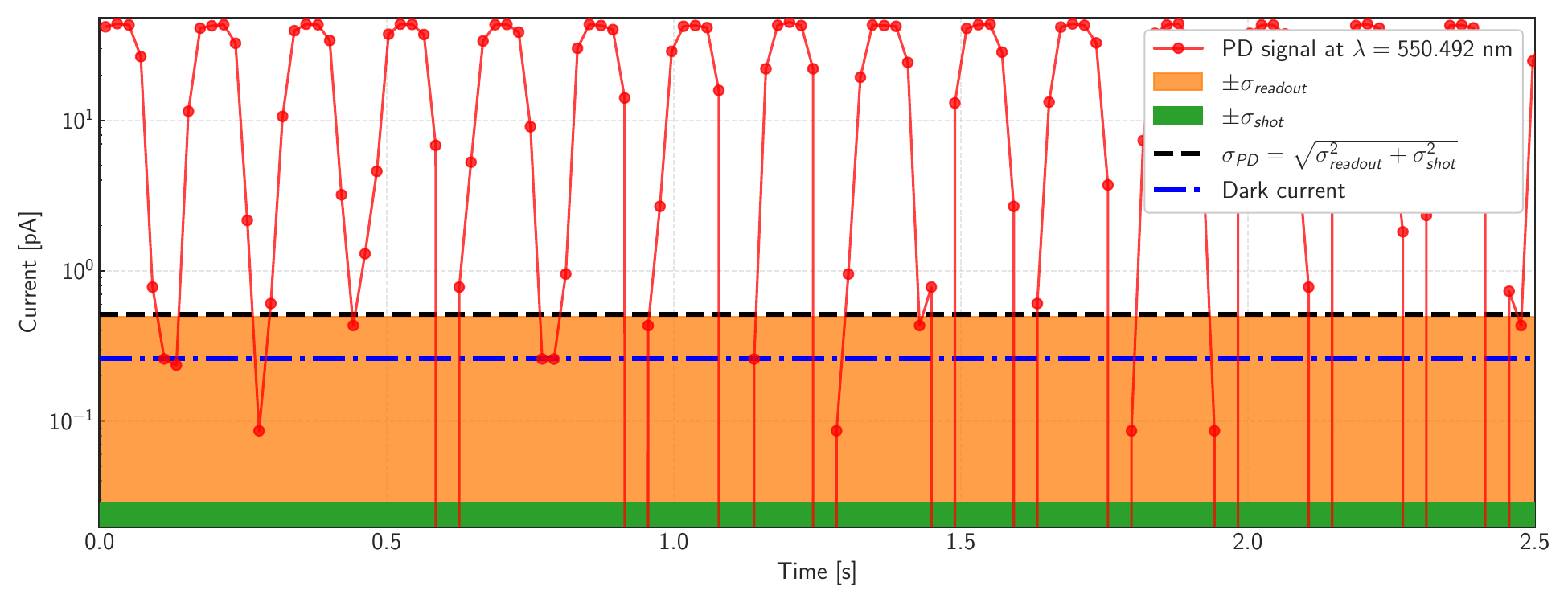}
  \caption{Semi-log scale illustration of the PD signal (red) superposed with different noise contributions. The orange area represents the total readout noise $\pm \sigma_{readout}$ obtained from the root-mean-square (RMS) of 4 independent dark signals. The green area is the shot noise $\pm \sigma_{shot}$. The blue line is the dark current $I_{dark}$. The chopped aspect of the signal is visible with a period of 1/6\,s set by twice the rotating mask frequency of 3\,Hz. The dashed black line represents the total noise $\sigma_{PD}$. {It should be noted that for the sake of visualization, we only display the maximum amplitude of the time-dependent noise time-series in the plot as horizontal lines.}}
  \label{fig:pd_raw_signal}
\end{figure*}

Moreover, despite having low impact on this experiment, the shot noise needs to be accounted for \citep{shotnoise}:
\begin{equation}
    \sigma_{shot}^{2}(\lambda, t) = 2e \times \left[I_{photocurrent}(\lambda, t) + I_{dark}\right] \times \Delta f
\end{equation}
with $\sigma_{shot}(t)$ the shot noise in A, $e$ the elementary charge of the electron $e = 1.602 \cdot 10^{-19}$ C, $\Delta f$ the bandwidth of measurement in Hz (i.e. the frequency range over which the noise is being considered to compute the noise power), $I_{photocurrent}(t)$ the photocurrent at time $t$ (i.e. the current produced by the photodetector due to incident photons), $I_{dark}$ the dark current (i.e. the current in the absence of light, often caused by thermal effects or other background sources and taken as the average of 4 dark exposures).
Therefore, the total PD photocurrent noise at time $t$ for the scanned wavelength $\lambda$ is {estimated as}
\begin{equation}\label{eq:pd_noise_model}
    \sigma_{PD}(\lambda, t) = \sqrt{\sigma_{readout}^{2} + \sigma_{shot}^{2}(\lambda, t)}
\end{equation}
Overall, photocurrent readings are dominated by the readout noise representing on average $\sim 85$ per cent of $\sigma_{PD}$. Fig.~\ref{fig:pd_raw_signal} shows the different noise contributions.

\subsection{Solar cell noise model}
\begin{figure*}
  \centering
  \includegraphics[width=1.0\linewidth]{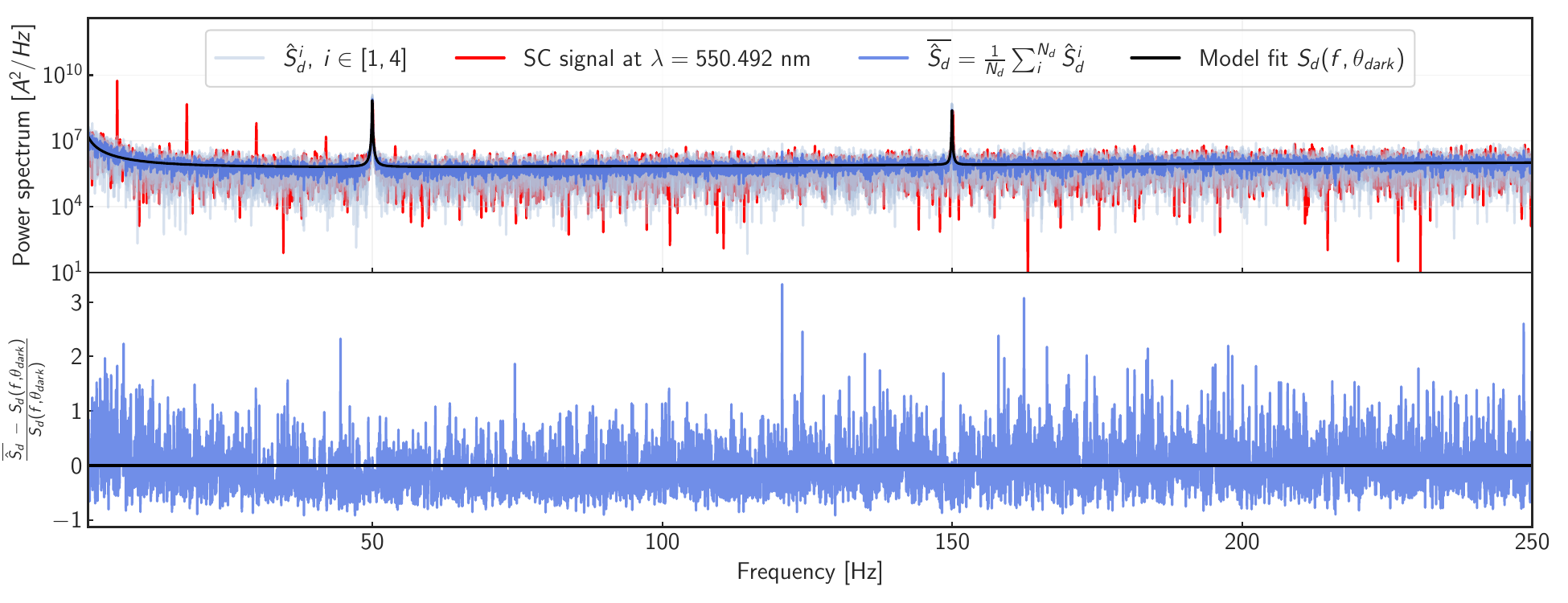}
  \caption{\textit{Top panel}: Power spectrum of the four SC dark signals (light gray), of their average (blue), and of the corresponding model best-fitting (black). For comparison, the signal at $\lambda$ = 550 nm is represented in red in the background. \textit{Bottom panel}: fractional residuals between the average dark power spectrum and the best-fitting model.}
  \label{fig:fit_sc_dark_signal}
\end{figure*}

The noise structure of the SC is much more complicated, as can be seen in Fourier space on Fig.~\ref{fig:fit_sc_dark_signal}: $1/f$ and electrical grid contributions
largely dominate third and higher orders amplitudes.
 
 We build a noise model for the SC (see equation~\ref{eq:sc_noise_model}), noted $\mathbf{{S}_{d}}(\mathbf{f}, \boldsymbol{\theta_{d}})$ with the vector of parameters $\boldsymbol{\theta_{d}} = \{\mathcal{A}, \beta, \mu, s\}$ (shown in black on Fig.~\ref{fig:fit_sc_dark_signal}) by fitting in Fourier space the averaged spectrum $\overline{\hat{S}_{d}}$ (shown in blue) of 4 SC darks (shown in gray). Hereafter, the overline notation specifies the average whereas the hat indicates an experimental measurement.
\begin{equation}\label{eq:sc_noise_model}
  \mathbf{S_{d}}(\mathbf{f}, \boldsymbol{\theta_{d}}) = \frac{\mathcal{A}}{\mathbf{f}^{\beta}} + \mu \times (1+s \times \mathbf{f}) + V_{50} + V_{150} \: ,
\end{equation}
\noindent where $\mathcal{A}$ and $\beta$ are the $1/f$ noise parameters, $\mathbf{f}$ is the vector of frequencies, $\mu$ and $s$ account for a small departure from white noise, and $V_{X}$ is a Voigt function modeling the 2 spikes at $X=50$\,Hz and $150$\,Hz: $V_{X}=a_{X}/(1+((\mathbf{f}-X)/w_{X})^{2})$ where $w_{X}$ is the peak width and $a_{X}$ is the peak amplitude. This model is fitted to the $\overline{\hat{S}_{d}}$ data, resulting in the error model subsequently used for the SC signal. As can be seen on the bottom panel of Fig.~\ref{fig:fit_sc_dark_signal}, the fit is reasonably good, considering the inherent difficulties in modeling all the noise contributions. Table~\ref{tab:sc_dark} shows the best-fit values with their associated 1-$\sigma$ uncertainties.

\begin{table}
	\centering
	\caption{SC dark signal best-fit values with their associated uncertainties and relative error. N/A denotes parameters directly extracted from experimental data using the tools from \textsc{scipy.signal}.}
	\label{tab:sc_dark}
	\begin{tabular}{l|c|c|r} % four columns, alignment for each
            \hline
            Parameter & Best-fit & $\pm$ 1-$\sigma$ & Rel. error [\%] \\
            \hline
            $\mu$ & $4.6311 \times 10^{5}$ & $6.3953 \times 10^{4}$ & 13.81 \\ [1.0 ex]
            $s$ & 0.0048 & 0.0015 & 32.07 \\ [1.0 ex]
            $\mathcal{A}$ & $1.5011 \times 10^{7}$ & $4.4387 \times 10^{5}$ & 2.96 \\ [1.0 ex]
            $\beta$ & 1.3105 & 0.0105 & 0.80 \\ [1.0 ex]
            $a_{50}$ & $4.3902 \times 10^{15}$ & $1.4595 \times 10^{13}$ & 0.33 \\ [1.0 ex]
            $f_{50}$ & 49.9800 & N/A & N/A \\ [1.0 ex]
            $w_{50}$ & $1.0000 \times 10^{-5}$ & N/A & N/A \\ [1.0 ex]
            $a_{150}$ & $1.5247 \times 10^{13}$ & $4.0815 \times 10^{10}$ & 0.27 \\ [1.0 ex]
            $f_{150}$ & 149.9400 & N/A & N/A \\ [1.0 ex]
            $w_{150}$ & $1.0000 \times 10^{-4}$ & N/A & N/A \\
            \hline
	\end{tabular}
\end{table}

\subsection{Photocurrent model}\label{sec:data_fitting}
As mentioned in Section~\ref{sec:light_source}, the chopper wheel shapes the intensity into a periodic signal of frequency $\nu$, which we model as follows. 
First, the time $t$ is folded into a phase $p\in[0,2[$,
\begin{equation}\label{eq:phase}
p = t \times \nu + \phi \equiv 0 \pmod 2
\end{equation}
\noindent where $\nu$ is the frequency or period of the periodic signal, and $\phi$ is a phase offset to centre the signal on the $[0,2[$ interval. The intensity received through one slit is then modeled as,
 \begin{eqnarray}
    h(p;\; \alpha, \tau_1, \tau_2) &=& \frac{g(p;\alpha, \tau_1, \tau_2) - g(0;\;\alpha, \tau_1, \tau_2)}{g(0.5;\;\alpha, \tau_1, \tau_2) - g(0;\;\alpha, \tau_1, \tau_2)} \text{\: with} \label{eq:gate}\\
 g(p;\;\alpha, \tau_1, \tau_2) &=& s(p;\; \alpha, \tau_1) - s(p;\;\alpha, 1-\tau_2) \text{\: and} \label{eq:raw_gate}\\ 
 s(p;\; \alpha, \tau) &=&  \frac{1}{1 + \exp\left(-\alpha \times (p-\tau)\right)}\: .\label{eq:sigmoid}
\end{eqnarray}
\noindent {Equation~\ref{eq:sigmoid} is the usual sigmoid function. 
In Equation~\ref{eq:raw_gate}, $\tau_2$ is defined so that its value is close to $\tau_1$. Equation~\ref{eq:gate} ensures that $h$ evaluates to 1 at $p=0.5$, and to 0 at $p=0$ and $p=1$.}
Finally, the modeled intensity is
\begin{align}
    I(p;\; A, \mu, \alpha, \tau_1, \tau_2) &=  \nonumber\\ 
     A\,[ h(p;\; \alpha, &\tau_1, \tau_2) \;+\; h(p-1;\; \alpha, \tau_2, \tau_1)]\;  +\; \mu
     \label{eq:improved_model}  
\end{align}
{where the first function accounts for the first slit of the chopper seen when a run starts ($p$ between 0 and 1) and the second function accounts for the second slit of the chopper ($p$ between 1 and 2) and has the parameters $\tau$ inverted. The parameter \textmu\ accounts for a fluctuating baseline.}
The amplitude $A$, in pA, is the main parameter of interest to determine.

\subsection{Signal fitting}
\begin{figure*}
  \centering
  \includegraphics[width=1.0\hsize]{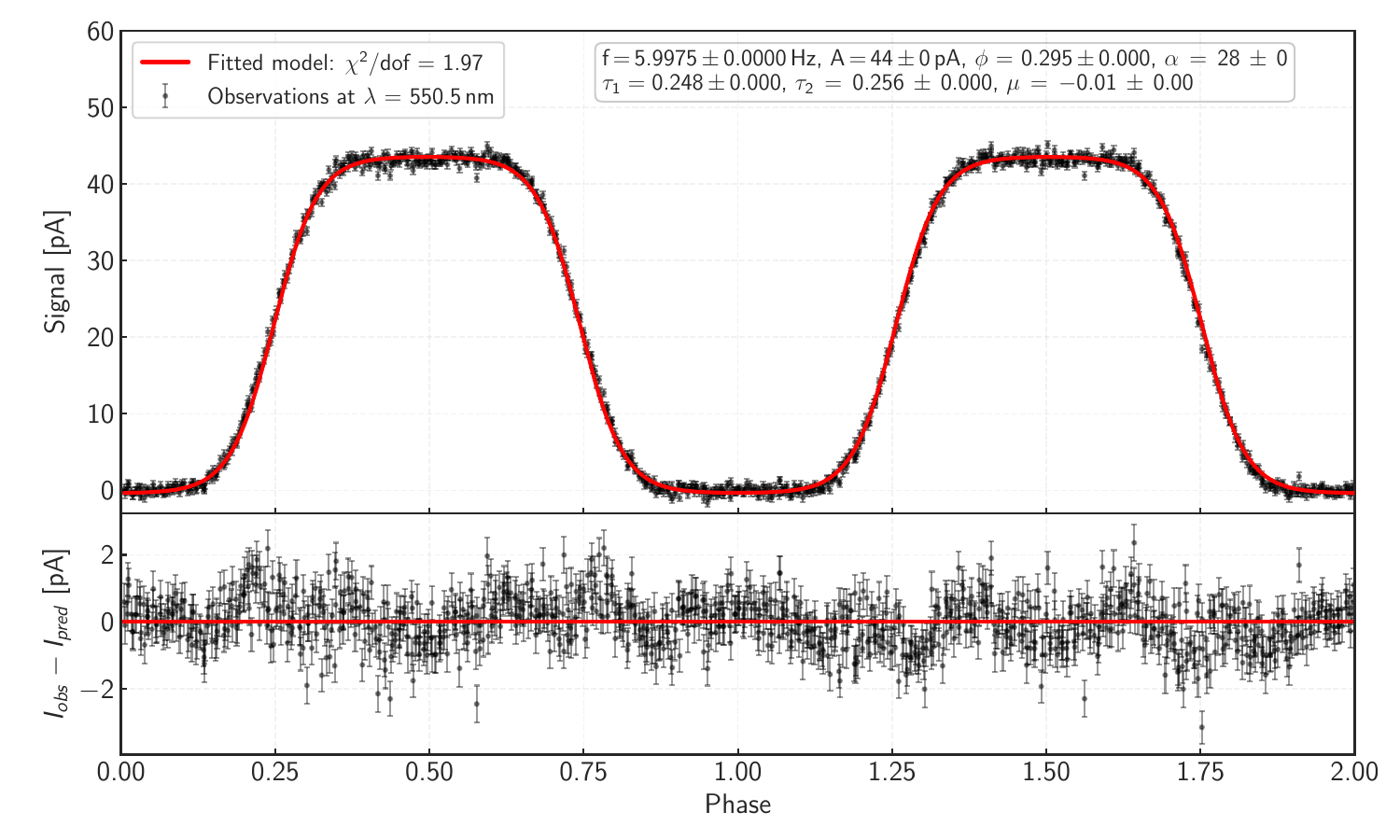}
  \caption{\textit{Top panel}: PD raw observations are plotted as dotted black points with $\pm$1-$\sigma$ uncertainty error bars and the red curve show the fitted model. Time series are folded between a single period. Best-fitting parameters with $\pm$1-$\sigma$ uncertainties are shown on the upper right. \textit{Bottom panel}: residuals between observations and best-fitting model. {They appear structured, showing some inadequacy of the model to catch small fluctuations of photocurrent caused by the blurred shape of the light beam.}}
  \label{fig:fit_double_slit_pd}
\end{figure*}

For each wavelength step, a $\chi^2$-minimization fit of the signal model is performed with \textsc{lmfit}\footnote{\url{https://lmfit.github.io/lmfit-py/}} on the PD and SC time-series, in time space for the PD and in Fourier space for the SC, using respectively Equations~\ref{eq:pd_noise_model} and \ref{eq:sc_noise_model} to define the covariance matrix used in the fit. Though it should be very close to 6\,Hz, the signal frequency $\nu$ is fitted to the data for better accuracy. For the PD, it appears that the baseline parameter $\mu$ shows a behavior correlated with the intensity of the light, which is not completely understood; as a result, we modify Equation~\ref{eq:improved_model} so that the amplitude A is factoring both the two $h$ functions and $\mu$.
Fig.~\ref{fig:fit_double_slit_pd} shows the PD data and fitted model for a wavelength corresponding to a SNR in the middle of the dynamic range. One can observe that the frequency is indeed very close to 6\,Hz, and that the residuals, though quite small (at most 6 per cent of the signal amplitude $A$), are structured in a way suggesting some imperfection related to the beam shape being blurred and having tails.

\begin{figure*}
  \centering
  \includegraphics[width=1.0\hsize]{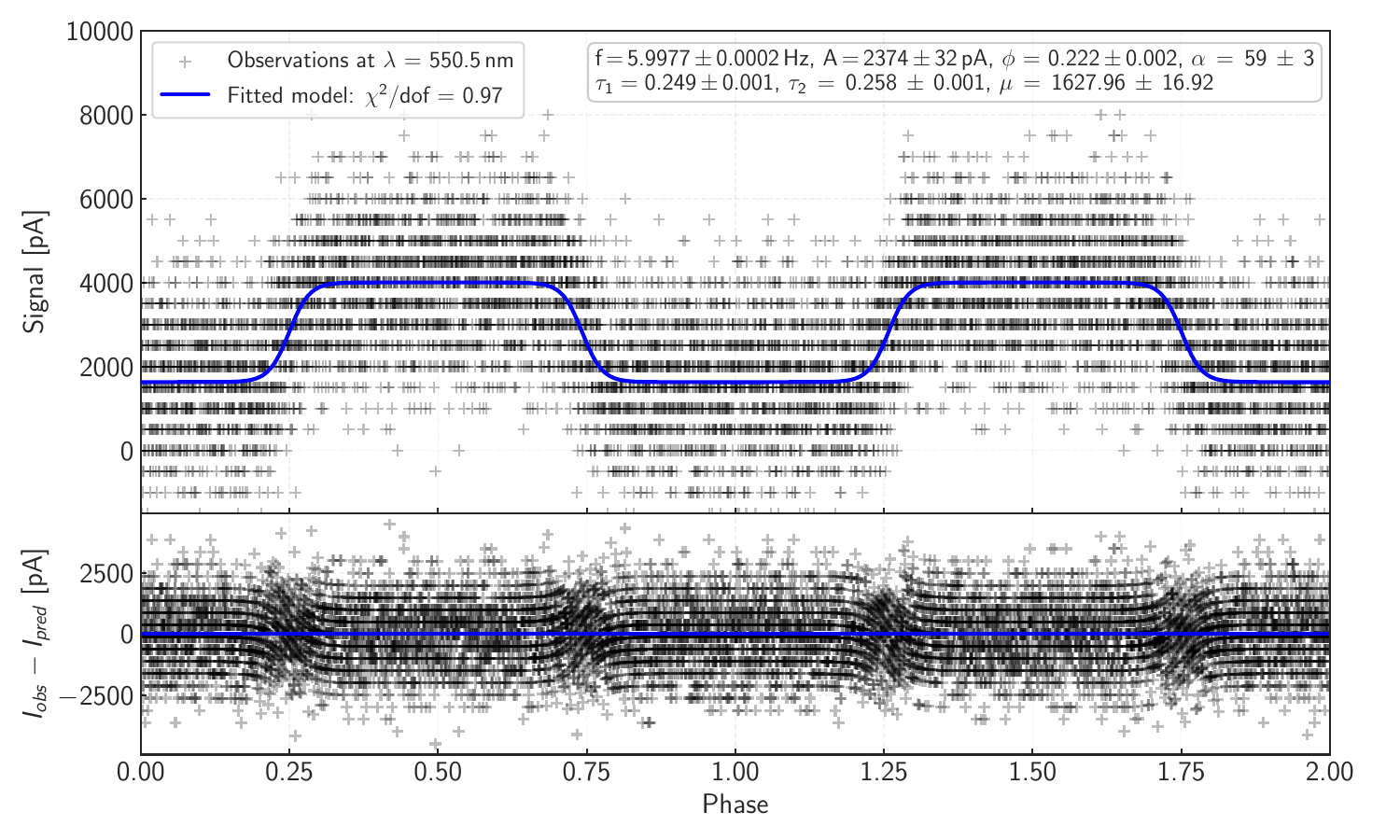}
  \caption{\textit{Top panel}: SC raw observations are plotted as black crosses and the blue curve depicts the fitted model. Best-fitting parameters with $\pm$1-$\sigma$ uncertainties are shown on the upper right.  \textit{Bottom panel}: residuals between observations and best-fitting model. {The presence of patterns in the residuals, observed during transition phases (when the photocurrent goes up and down at points 0.25, 0.75, 1.25, and 1.75), is believed to be caused by oscillations in a 50 Hz signal.}}
  \label{fig:fit_double_slit_sc}
\end{figure*}

Fig.~\ref{fig:fit_double_slit_sc} shows the SC data and fit result to observations at a wavelength with medium SNR. The much degraded level of noise in time space, when compared to the PD, is apparent, together with the digitization of the Keysight's readings.

\section{Results}
\label{sec:results}

\begin{figure*}
  \centering
  \includegraphics[width=\hsize]{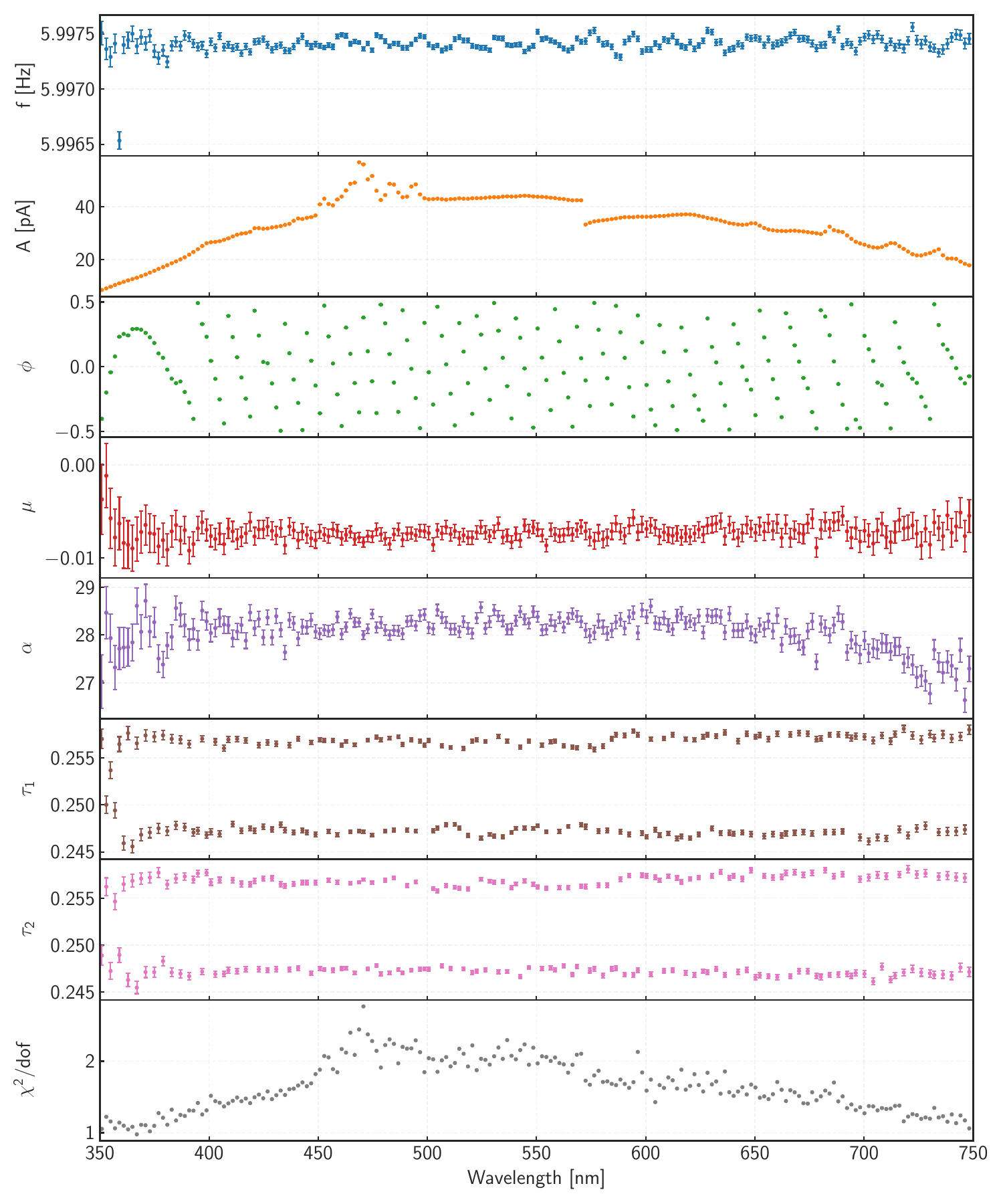}
  \caption{Results for modeling of the full scan PD data. From top to bottom : frequency $f$, amplitude $A$, phase $\phi$, offset $\mu$, slope $\alpha$, slits widths $\tau_{1}$ and $\tau_{2}$, and $\chi^{2} / dof$. Bi-modality of $\tau_{1}$ and $\tau_{2}$ parameters originates from the chopper slits. The linear decreasing trend of the phase offset parameter $\phi$ is thought to be a constant time lag between consecutive observations at high SNR. The settling time is somewhat constant with a small gap increase caused by the movement of the monochromator's grating turret performed to adjust the grating position that produces the desired wavelength. After reaching the fit lower boundary limit of -0.5, it reaches again a value close to the upper limit of 0.5.}
  \label{fig:final_results_pd}
\end{figure*}

\begin{figure*}
  \centering
  \includegraphics[width=\hsize]{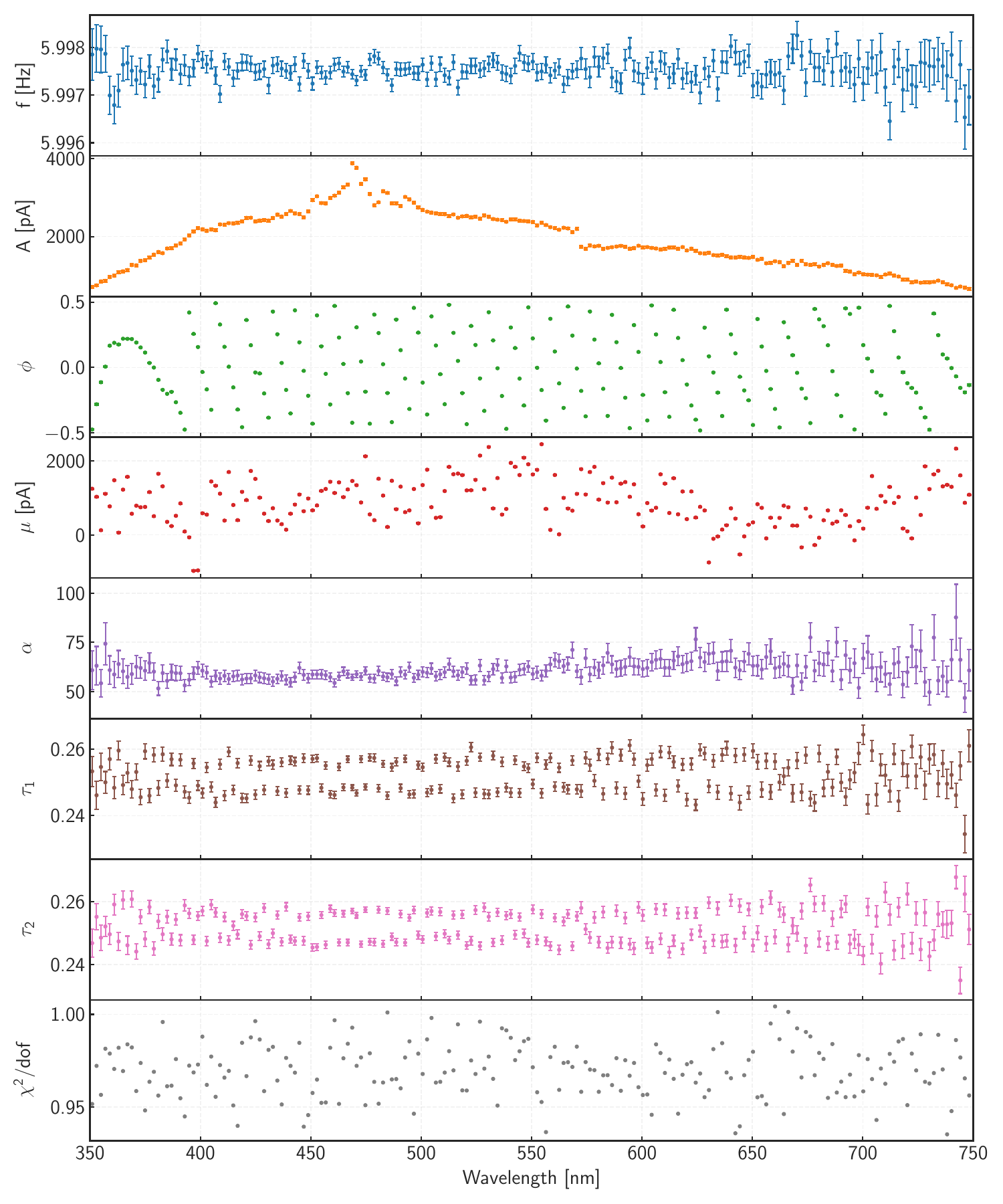}
  \caption{Same as Fig.~\ref{fig:final_results_pd} for the SC data.}
  \label{fig:final_results_sc}
\end{figure*}

Performing the fits to the PD and SC signals at each wavelength yields the results presented in Fig.~\ref{fig:final_results_pd} and \ref{fig:final_results_sc}, respectively. {Some comments are in order}. First, the wavelength dependence of the signal amplitudes mostly follows that of the light source. The discontinuity at 570\,nm is the result of the addition of the high-pass filter discussed in Section~\ref{sec:light_source}. Although being without consequences due to the error budget being largely dominated by the SC (see Section~\ref{subsec:model_error}), the fact that the reduced $\chi^2$ for the PD follows the same trend is indicative of an incomplete error modelization.

We can now proceed with the main results of this work. We first determine the relative throughput curve of our setup. Then, to assess its performance for ZTF photometry, we explore the effect of uncertainties on the zero-point $zp$ and the central wavelength $\overline{\lambda}$ of the ztf-g and ztf-r filters.

\subsection{Determination of the CBP telescope throughput}
\begin{figure*}
  \centering
  \includegraphics[width=1.0\linewidth]{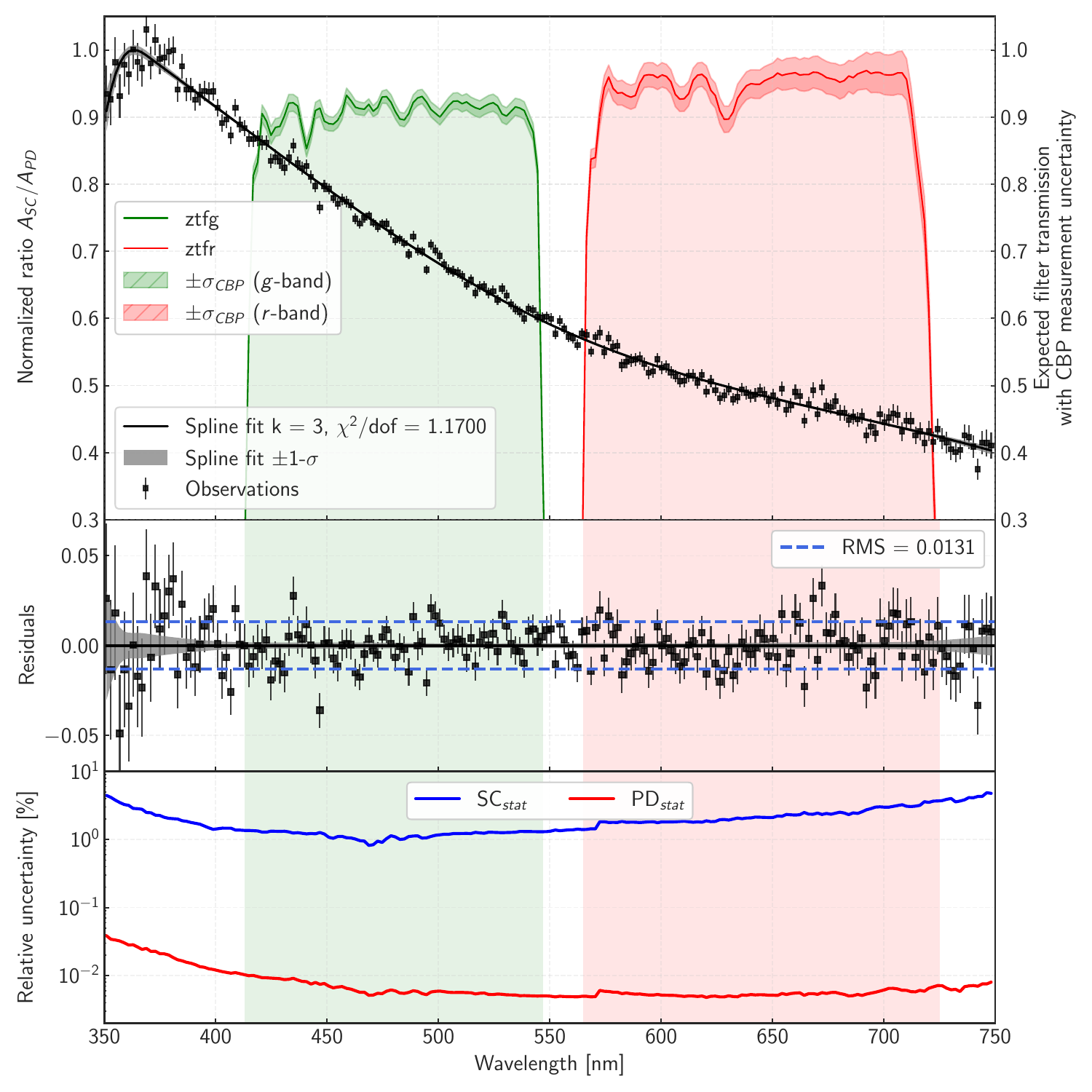}
  \caption{\textit{Top panel}: Ratio of the photocurrent amplitudes $A_{SC}/A_{PD}$ as a function of wavelength. Black error bars depict individual measurements at each scanned wavelength with $\pm$1-$\sigma$ statistical uncertainty. The solid black line is a cubic spline fit to the data, with a gray shaded area showing the $\pm$1-$\sigma$ uncertainty of the fit, assuming the telescope throughput and the quantum efficiencies are smooth as shown by previous higher-resolution measurements. Theoretical filter transmissions are represented with a $\pm$1-$\sigma$ uncertainty corresponding to the CBP throughput measurement uncertainty. \textit{Centre panel}: residuals between the spline fit and observations. Blue dashed lines are drawn at the RMS level. \textit{Bottom panel}: Contributions to the total uncertainty are shown for the SC amplitude determination (blue curve) and the PD amplitude determination (red). The total statistic error is not shown as it overlaps with the SC uncertainty contribution curve. The gradual rises on the sides are due to lower light source power (near UV) and lower grating efficiency (near IR).}
  \label{fig:CBP_response}
\end{figure*}

The QE of the SC was measured against a NIST-calibrated photodiode, to 0.1 per cent relative uncertainty \citep{Brownsberger2022}. It therefore measures accurately {the number of photons} emerging from the CBP. {By comparing the ratio of the SC photocurrent to the PD current, we can use the latter as a proxy to retrieve photon flux on other telescopes and determine their throughput\footnote{The absolute throughput $T_{CBP}(\lambda)$ of the CBP telescope can be calculated by introducing the quantum efficiencies of the PD and SC, respectively noted $\eta_{PD}$ and $\eta_{SC}$. It is defined as $T_{CBP}(\lambda) = \frac{A_{SC}(\lambda)}{A_{PD}(\lambda)} \times \frac{\eta_{PD}(\lambda)}{\eta_{SC}(\lambda)}$}}.
Hereafter, we consider the normalized ratio $A_{SC}/A_{PD}$ depicted on Fig.~\ref{fig:CBP_response}.
It is assumed that all flux loss between devices is due to the telescope itself (i.e. mirrors with coating, internal baffling).
The total {statistical} uncertainty on the CBP telescope throughput measurement is expressed as,
\begin{equation}
	\sigma_{CBP}^{2}(\lambda) = \left(\frac{A_{SC}(\lambda)}{A_{PD}(\lambda)}\right)^{2} \times \left[ \left(\frac{\sigma_{A_{SC}}(\lambda)}{A_{SC}(\lambda)}\right)^{2}+\left(\frac{\sigma_{A_{PD}}(\lambda)}{A_{PD}(\lambda)}\right)^{2} \right]\: .
\end{equation}
$\sigma_{A_{SC}}(\lambda)$ and $\sigma_{A_{PD}}(\lambda)$ arise directly from the weighted least-squares fit presented in Section \ref{sec:data}.

For the $g$ filter (see spectral range illustrated as a green area in Fig.~\ref{fig:CBP_response}), we have an average statistical uncertainty close to $\sim$~1 per cent for each data point. It increases to $\sim$~2 per cent for the $r$ filter (red area in Fig.~\ref{fig:CBP_response}). SC noise overwhelmingly dominates the total statistical uncertainty budget. The PD contribution to the statistical uncertainty is $\sim$ 0.017 per cent relative uncertainty. The increase seen near 350\,nm is due to the decrease in light source brightness. In contrast, the increase seen around 600\,nm is due to the low monochromator grating efficiency. \citet{Souverin2022} showed with higher-resolution measurements that the telescope transmission is smooth, as are the quantum efficiencies of the PD and SC. We can use this behavior to interpolate the data with a smooth function, reducing the global statistical error on the response curve. The solid black line is a cubic spline fit. Residuals between the observations and the spline interpolation are shown in Fig.~\ref{fig:CBP_response} and no structure seems to appear. It would be valuable to conduct repeated measurements to examine the consistency of observations around the fitted smooth spline curve and improve the fit quality.

\subsection{Errors on magnitude estimation}
\label{sec:zp_err}
We can estimate the impact of the uncertainty of the CBP throughput measurement by calculating its effect on photometric measurements. We first compute the effect of the uncertainty on the filter transmission curve on magnitudes.
The magnitude $m_{X}$ of a flat spectrum measured with a photon counting device through a filter $X$ of transmission $T_X(\nu)$ is
\begin{equation} 
     m_{X} = -2.5 \times \log_{10}\left(\frac{\int T_{X}(\nu) \cdot (h \nu)^{-1} d\nu}{\int (h \nu)^{-1} d\nu}\right),
     \label{eq:mag}
\end{equation}
with $h$ the Planck constant and $\nu$ the photon frequency.
To evaluate the effect of the CBP throughput error, we draw random samples from a normal distribution $\mathcal{N}(\mu = T_{X}(\lambda_{i}), \: \sigma = \sigma_{CBP}(\lambda_{i}))$ that we inject onto the magnitude equation, and calculate the mean and standard deviation $\sigma_{zp, X}$ of the magnitude distribution we obtain.
We propagate uncertainties on the magnitude zero-point for a flat spectrum. For the ztf-g and ztf-r bands respectively, we obtain similar zero-point statistical uncertainties with $\sigma_{zp,g} = 0.0049$ mag and $\sigma_{zp,r} = 0.0052$ mag.

\subsection{Central wavelength uncertainty}
\label{sec:lambda_err}
The second key metric is the accuracy of the bandpass mean wavelength, i.e. the central wavelength $\overline{\lambda}$ for a flat spectrum:
\begin{equation}
    \overline{\lambda_{X}} = \frac{\int \lambda \cdot T_{X}(\lambda) \cdot d\lambda}{\int T_{X}(\lambda) \cdot d\lambda},
    \label{eq:central_wavelength}
\end{equation}
 Here, estimations are done in the same manner as for the zero-point magnitude uncertainty estimation $\sigma_{zp,X}$ by drawing random samples from a normal distribution. Both filters appear determined with sub-nanometer accuracy. For the ztf-g and ztf-r bands respectively, we obtain $\sigma_{\overline{\lambda}_{g}} = 0.9$\,nm and $\sigma_{\overline{\lambda}_{r}} = 0.6$\,nm. As the filters have large bandpasses $\Delta \lambda_{X} = \lambda_{max,X} - \lambda_{min,X}$ (135 nm and 160 nm for ztf-g and ztf-r filters respectively), relative uncertainties $\sigma_{\overline{\lambda}_{X}}/\Delta \lambda_{X}$ of central wavelengths are less than 0.5 per cent for both filters.

\section{Evaluation of uncertainties}
\label{sec:measurement_uncertainty}

In our efforts to uphold an accurate uncertainty budget for the CBP system, we have made diligent attempts to address the potential effect of the model error. In this section, we establish an assessment of the uncertainty budget and explore three sources of systematic uncertainty that could impact our results.

\subsection{Overview of potential systematic uncertainties}

As our primary focus was to demonstrate the feasibility of the concept itself through the realization of this non-definitive prototype, we did not conduct an exhaustive qualitative and quantitative evaluation of each systematic effect. However, it is essential to recognize that these factors may warrant more in-depth investigation in future iterations of this work to refine the accuracy of our measurements. Below, we detail two main systematic effects affecting this particular experimental setup.

\textbf{Ambient and stray light contamination} : the monochromator manufacturer datasheet\footnote{\url{https://www.newport.com/medias/sys_master/images/images/h2c/hc0/9134874230814/DS-121402-Cornerstone-130.pdf}} indicates a 0.03 per cent stray light level. The longpass filter manufacturer datasheet\footnote{\url{https://www.thorlabs.com/newgrouppage9.cfm?objectgroup_id=918}} gives an optical density (OD) in excess of 4, corresponding to 0.01 per cent transmission in the rejection spectral region. The presence of electronic components in the room (electrometers, PC monitors, and computer status LEDs), while conducting the full wavelength scan, may have impacted the beam characteristics, particularly because the IS was fully open at the beam entrance port. Furthermore, the insulation of the Xenon lamp was not perfect and some light leaked into the room. Careful shielding and light-blocking measures may be necessary to mitigate this effect.

\textbf{Photodiode temporal and thermal stability}: previous work by \citet{Eppeldauer90} for the optical spectrum range (400-750\,nm) showed that the spectral response changes of standard silicon photodiodes were as high as 0.5 per cent / year at constant operating temperature. Based on \citet{Larason2008} findings, the temperature-dependent change becomes significant only for wavelengths exceeding 1000\,nm, where it corresponds to 0.2 per cent/°C. 
\citet{Stubbs2010} are more conservative with an estimate of 1 per cent/°C above 900 nm and 0.1 per cent/°C for $\lambda <$ 900\,nm. Continuous monitoring of the photodiode's temperature may be required to ensure reliable and accurate data acquisition.

\subsection{Model error}
\label{subsec:model_error}

\begin{figure*}
  \centering
  \includegraphics[width=\hsize]{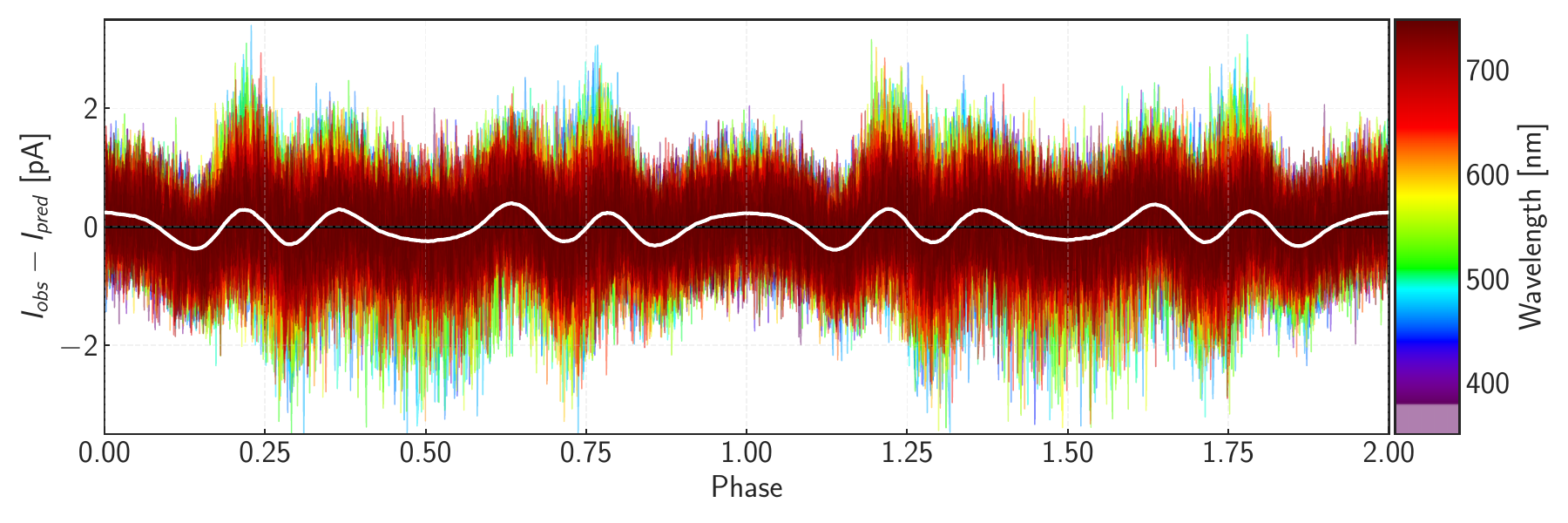}
  \caption{Residuals between observations and best-fitting model for the PD signal for all wavelengths (with coherent matching colors). The asymmetric pattern appears strongly defined. The mean curve of residuals is plotted in white.}
  \label{fig:pd_residuals}
\end{figure*}

The model is an accurate description of the dataset except for specific features in the residuals $\boldsymbol{\Delta I} = \boldsymbol{I_{obs}} - \boldsymbol{I_{pred}}$ that we call "model structured error" and is seen at each observation wavelength for the PD signal (see Fig.~\ref{fig:pd_residuals}). We suppose these oscillations to be caused by the monochromatic light beam shape being blurred at the entry of the chopper wheel. If the shape has tails, the chopping mechanism may produce such small photocurrent oscillations and does not conform to a perfect sigmoid curve.
The goal is to check if these periodic oscillations downgrade the fit quality if they impact the determination of the amplitude parameter $A_{PD}$ or add any bias. For SC signals, the residual structure is below the noise level.\\

\noindent Two kinds of simulations are undertaken :
\begin{enumerate}
    \item \textbf{Measurement noise only}. For the photodiode, we have direct access to a good estimate of the uncertainty $\sigma_{PD}$. We simulate a signal with the fit parameters $\theta_{PD}(\lambda_{i})$. We draw random samples from a multivariate normal distribution using the model parameters covariance matrix obtained from the linear regression. Only the amplitude is kept at the original fit value. On these "perfectly" simulated signals, we add a random noise from a normal distribution $\mathcal{N}(0, \sigma_{PD})$. The signal is then binned at the multimeter resolution of 0.17\,pA. For the solar cell, we simulate the correlated noise by contaminating a perfectly simulated signal with 50\,Hz and 150\,Hz lines and 1/f$^{\beta}$ noise directly in the time series. We similarly bin points at the multimeter resolution of 500\,pA.
    \item \textbf{Measurement noise + modeling errors in the shape of the signal}. Residuals between the model and PD observations are interpolated at each wavelength step $\lambda_{i}$. Fig.~\ref{fig:pd_residuals} depicts the shape of these residuals. These are added to the same theoretically perfect simulated signals and standard measurement errors are added afterward.
\end{enumerate}
At each wavelength and for the two simulated cases, we sample and fit afterward $N_{sim} = 10^{4}$ SC and PD signals. The reconstructed parameters provide information on the effect of potential systematic effects introduced or omitted by the model. \ref{fig:contours_pd_sim_dual} and \ref{fig:contours_simulation_sc} give an example of 2-$\sigma$ contours obtained at the dataset median wavelength $\lambda = 550$\,nm for both simulation cases.
Overall, we find good agreement between model observations and simulations. No significant deviations between simulations and observations appear. 
We find a small constant bias of $\sim$0.003\,pA on all the simulated PD signals when adding the structured residuals in simulation n°2. This is within the uncertainty of our $A_{PD}$ estimate {and below the electrometer resolution}, so the shift can be neglected. The average statistical uncertainty on $A_{PD}$ stands at $\sim$ 0.05\,pA for the measurement noise simulation and $\sim$ 0.09\,pA for the simulation with the added residuals. The model uncertainty contributions follow the expression
\begin{equation}
     \sigma_{model, \: A_{PD}}^{2} = \sigma_{meas+res, \: A_{PD}}^{2} - \sigma_{meas, \: A_{PD}}^{2}
\end{equation}
$\sigma_{model, \: A_{PD}}^{2}$ represents the uncertainty contribution of the model, $\sigma_{meas+res, \: A_{PD}}^{2}$ represents the total uncertainty (simulation n°2) and $\sigma_{meas, \: A_{PD}}^{2}$ represents the measurement uncertainty only (simulation n°1). The non-modeled residuals structure contributes to $\sim$ 83 per cent of the variance in the $A_{PD}$ estimate and the remaining 17 per cent originates from the intrinsic measurement noise.

\begin{figure}
  \centering
  \includegraphics[width=\columnwidth]{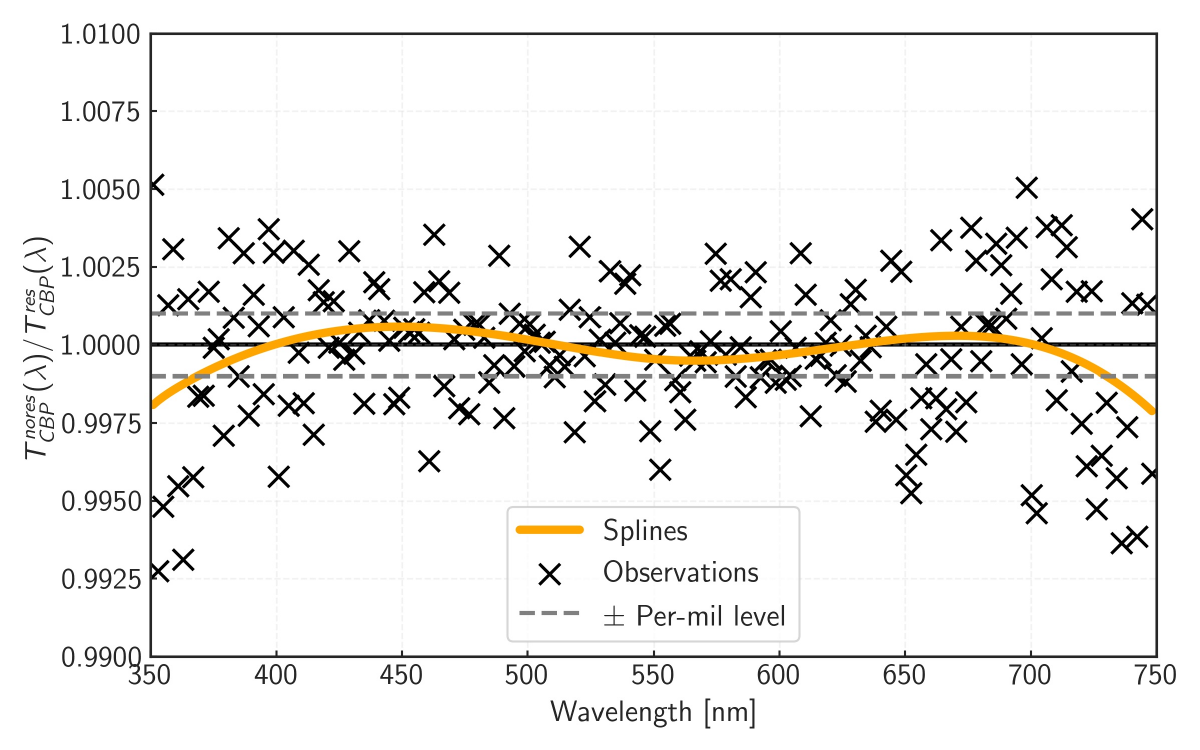}
  \caption{Ratio of CBP throughput curves for fitting PD signal with and without residuals (see Fig.~\ref{fig:pd_residuals}). Individual samples are plotted in black crosses and the spline fit ratio is drawn in orange curve. No bias is visible. Deviations at low and high wavelengths are due to SNR decreasing.}
  \label{fig:ratio_models_sim}
\end{figure}

{However, the PD accounts for less than 0.05 per cent of relative uncertainty to the total CBP throughput uncertainty budget}. The error budget is completely dominated by the SC signal noise. Thus, our model fitting observations without integrating the "model structured error" contribution is sufficiently good as long as the SNR on the SC signal does not drastically increase. Ratios of CBP throughput curves for the considered simulated cases are shown in Fig.~\ref{fig:ratio_models_sim}. No wavelength dependency appears and the ratio of splines stays below the per-mil level.

\section{Discussion}
\label{sec:discussion}

As seen before, the current prototype of the traveling version of the CBP looks very promising. In order to be usable in the context of UV-VIS-NIR photometric survey instruments, some improvements on both the equipment and the acquisition method are required. We discuss their potential implementation hereafter.

\subsection{Reducing statistical uncertainty}

\subsubsection{Upgrading the light source to increase photon flux}
Integration time is one of the key constraints to the design and performance of a CBP, together with the overall stability and precision. Indeed, telescopes operate for whole nights. That makes them available for calibration purposes during the day or more if the weather prevents observations. Moreover, integration times need to be compatible at the millimagnitude level for potential interference from daylight, given that telescope domes do not provide complete light insulation. Therefore, reaching the desired level of uncertainty on the instrument curve throughput determination requires integration times as short as possible. With a 175W Xenon light as we had for this preliminary feasibility test, $t_{int}=20$\,s is insufficient to reach the per-mil level of {statistical} uncertainty. For that reason, a more powerful light source must be found. Various options exist, such as laser-driven light sources (LDLS) which benefit from laser power without the large volume occupancy or potential hazard. 
They can also have a broad wavelength range (from UV to IR) with very high spectral radiance (tiny plasma diameters up to 1 mm producing radiance ranging from 1 to 100\,mW/mm$^{2}$/sr/nm, much higher than the radiance in our setup of $\sim$ 0.05\,mW/mm$^{2}$/sr/nm at peak), stability and lifetime. 
Furthermore, collimating light from traditional sources often results in a beam divergence or widening angle, which can be problematic. The emission surface is significantly higher for arc-lamps with millimeter-sized ones against $\sim$ 300 \textmu m for the LDLS. The small emitting point of LDLS allows for efficient focusing of the light onto a very small area.
Replacing the 175W Xenon light source with the IDIL TLS3-EQ77-UV-VIS-NIR monochromator and LDLS bundle is considered. The coupling between these two parts is achieved with optic fibers.
Statistical uncertainty on each wavelength data point can be decreased by one or even two orders of magnitude if the effective radiance entering the IS is proportionally higher. Nevertheless, it is important to stay in tune with the upper flux limit imposed by the telescope to calibrate for the real-use case scenario of the traveling CBP instrument. This issue arises due to the potential saturation of the CCD/CMOS camera, resulting from the injection of an excessive amount of monochromatic flux into the sensor within a brief exposure time.

\subsubsection{Speeding up the chopping frequency}

One caveat observed in post-processing is that the signal frequency set at 6 Hz is too slow to avoid contamination by the 1/f noise. To quantify the potential gain expected from running at higher {chopping} frequency, we run simulations using our models. Knowing the fitted parameters of each signal for both SC and PD at each wavelength of the acquisition, we increase the frequency to $f_{2}$ = 12\,Hz and $f_{3}$ = 18\,Hz to evaluate the impact on the estimated amplitudes. Signal-to-noise ratio improve by $\sim$ 25 per cent when going from 6 to 12\,Hz and $\sim$ 50 per cent from 6 to 18\,Hz. Above 18\,Hz, the overall gain of SNR becomes smaller because the signal peak is already far from the 1/f curve. Thus, rotating the chopper wheel motor at twice or three times the initial frequency or designing a mask with two additional symmetric holes will lead to a noticeable improvement of the SNR on the CBP throughput curve. {For the chopping system, off-the-shelf options are plentiful, or even careful redesign to improve our custom DIY device is feasible}.

\subsection{Mitigating systematic uncertainties}

\subsubsection{Monitoring sensors temperatures}
Accurate and reliable measurements with a CBP require careful monitoring of the temperature of critical components such as the photodiode and solar cell. The responsivity of these components is known to be sensitive to temperature variations, especially in the infrared $\lambda >$ 1000\,nm where responsitivity deviations can reach 1 per cent/°C for silicon-based photodiodes.

\subsubsection{Double monochromator configuration}

Achieving both the desired spectral radiance and bandwidth is difficult with a monochromator and a broadband light source as demonstrated in \citet{Stubbs2010}. In the past, a tunable laser was used to overcome this issue for previous CBP versions. In this paper, we have seen that target accuracies are within reach despite using an inappropriate monochromator. Our monochromator model was optimized for particle physics experiments (i.e. Cerenkov effect measurements, explaining the 250 nm blaze wavelength of the unique grating installed). It suffered from poor efficiency in the exploited range of observations averaging at 40 per cent for $\lambda$ = 550 nm (see Fig.~\ref{fig:spectrum_throughputs}). The etendue is also low as the grating width $W_{g}$ = 30 mm. As it is squared in the etendue equation, we would benefit tremendously from a grating twice as large (factor 4 of improvement in terms of flux). Upgrading the monochromator with appropriate larger gratings (UV + VIS + IR) could drastically increase the flux entering the IS, the SNR and reduce the integration time all at once. For example, gratings up to 68 mm widths are available for purchase with $>70$ per cent efficiency at peak. With this kind of configuration, an increase of around one order of magnitude on the signal-to-noise ratio can be expected at every wavelength sample.
Nonetheless, two potential additional systematic errors to address arise from the spectral bandwidth of the beam. While keeping the necessary amount of flux inside the IS to know the system transmission at better than 0.1 per cent level, it is also important to get beam spectra as narrow as possible in order to adequately capture the steep edges of filters ($\sim$ 10 nm for ZTF filters) with a high spectral resolution. The recommendation stands to have optical bandwidths of $\sim$ 1 nm. The second systematic effect to consider involves the variation in the spectral bandwidth shape with wavelength. In Fig.~\ref{fig:spectra_350}, we observe the wavelength dependency of the FWHM of the monochromatic beam generated by the monochromator. The FWHM decreases from $\sim$ 3.5 nm at $\lambda$ = 350 nm to $\sim$ 1.8 nm at $\lambda$ = 750 nm. This change impacts the measurement of the slope of filter edges since the observed CBP telescope response function is the convolution of the monochromatic filter response with the spectral bandwidth of the monochromator's beam. The width evolution leads to systematic and significant differences at the red and blue edges of a filter, causing a tendency to shift the observed filter curve in one direction or another.
Adding a second monochromator can tackle multiple sources of systematic uncertainties at once by reducing ambient light contamination (with a fully fibered optical path), enhancing spectral purity (light beam is diffracted one additional time), reducing out-of-band emission, and decreasing the beam spectral bandwidth. Loss of light flux is to be anticipated but may be compensated by carefully injecting light into the entrance slit of the first monochromator, and with twice as large and efficient gratings.

\subsection{Effect of individual sample uncertainty and wavelength scan step on zero-point and central wavelength precision}

\begin{figure}
  \centering
  \includegraphics[width=\columnwidth]{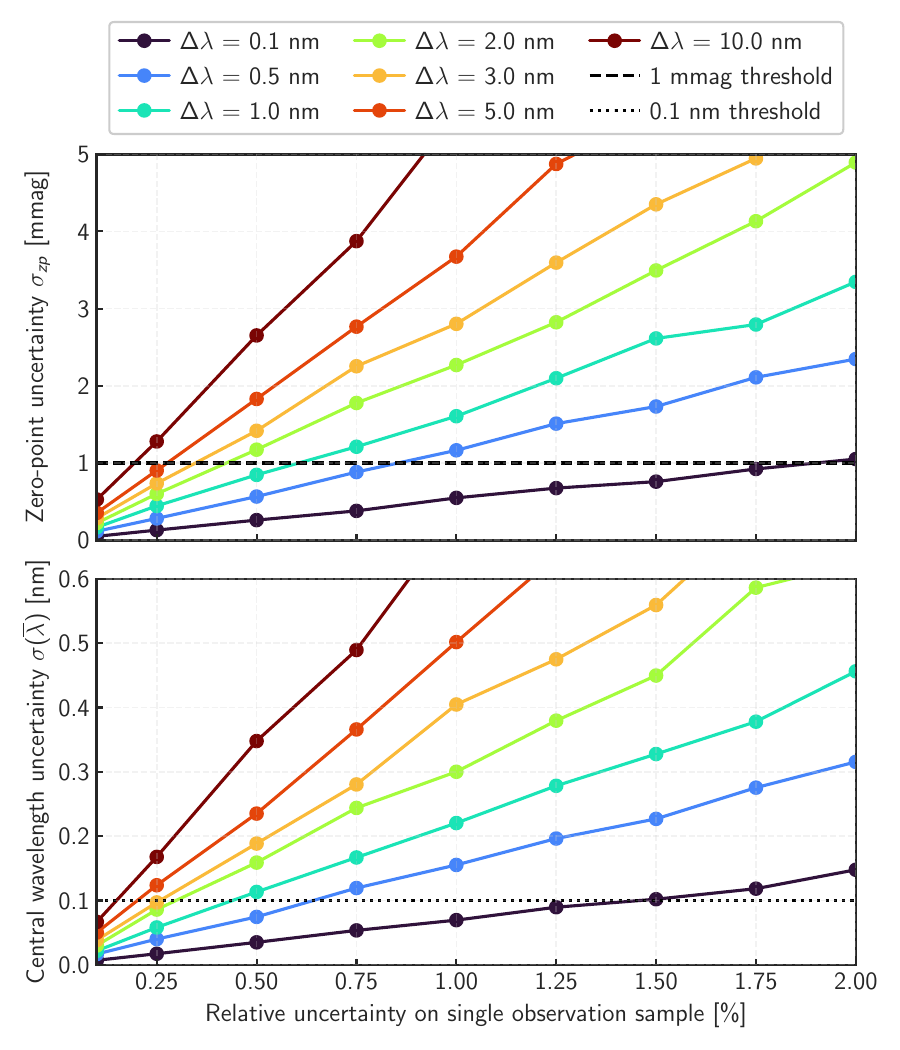}
  \caption{\textit{Top panel:} forecasts of zero-point magnitude uncertainty $\sigma_{zp}$ as a function of individual observation sample relative uncertainty $\sigma_{CBP}(\lambda)/T_{CBP}(\lambda)$ [per cent] and scan step $\Delta \lambda$ [nm] for the ztf-g band. \textit{Bottom panel:} same for central wavelength uncertainty $\sigma(\overline{\lambda})$ forecasts. Similar results are found for the ztf-r filter.}
  \label{fig:metrics_forecast}
\end{figure}

In planning the next iteration of the traveling CBP instrument, we present forecasts for statistical uncertainties in zero-point magnitude $\sigma_{zp}$ and central wavelength $\sigma_{\overline{\lambda}}$ metrics. This assessment serves two purposes: (i) to determine the necessary level of precision for individual observation samples, and (ii) to establish the required wavelength scan step ($\Delta \lambda$ in nm). Fig.~\ref{fig:metrics_forecast} depicts the anticipated uncertainty levels in zero-point and central wavelength for the ztf-g band. {We proceed in the same way as in Sections~\ref{sec:zp_err} and \ref{sec:lambda_err} but with a single relative uncertainty value for all samples. This relies on the assumption of a flat relative uncertainty distribution across the wavelength range. Then, we evaluate the metrics $\sigma_{zp}$ and $\sigma_{\overline{\lambda}}$ with Equations~\ref{eq:mag} and \ref{eq:central_wavelength} for different wavelength steps $\Delta \lambda$.}
Maintaining a constant wavelength scan step of $\Delta \lambda$ = 2\,nm, as chosen in this study, will allow reaching the mmag threshold in zero-point accuracy and a 0.1\,nm threshold in central wavelength accuracy, provided that the uncertainty in individual observation samples for the CBP telescope throughput curve is reduced to approximately 0.4 per cent. 
This corresponds to a factor 4 to 6 improvement considering ztf-g and ztf-r bands respectively, from the current prototype setup. 
Consequently, it is important to give careful consideration to the balance between the overall count of observation samples (which directly affects the total integration time for the CBP throughput scan) and the SNR of each observation sample (requiring applicable solutions to mitigate the SC noise). We emphasize that future work should address this trade-off and explore solutions to minimize exposure time (preventing sensor sensitivity variations due to changes in environmental conditions over large periods of time) and maximize SNR in the SC.

\section{Conclusions}
\label{sec:conclusions}

We assessed the replacement of a laser with a broadband light source to design a traveling Collimated Beam Projector. The critical aspect of designing such a calibration device is the measurement accuracy of the intermediary telescope acting as a light projector. By calibrating the CBP telescope relative to a solar cell itself calibrated at better than 0.1 per cent onto the NIST reference photodiode, we can propagate the calibration to telescope systems. The proof-of-concept setup presented in this paper has shown promising results for its intended use. It delivers average statistical uncertainty below 2 per cent in the 350 to 750 nm range which propagates into zero-point magnitudes uncertainties of $\sim$ 5 mmag for ztf-g and ztf-r bandpasses. {Filters central wavelengths would be determined at the sub-nanometer precision, considering statistical uncertainties only}. In its current state, the accuracy of the throughput curve is limited by the solar cell intrinsic dark noise accounting for nearly the entire statistical uncertainty budget. {The SC still remains the preferred intermediary calibration sensor because it possesses a large photosensitive surface area adequate for capturing the entirety of photons emitted through the CBP telescope.} To alleviate it, multiple efforts of methods and equipment improvements must be undertaken : (i) increasing the photon flux in the IS by using wider and more efficient gratings, upgrading the light source and refining the light injection mechanism; (ii) chopping the light beam at two or three times the current frequency to diminish the contamination of 1/f noise on the SC measurements; (iii) multiple scans as statistical uncertainty is expected to decrease by $1/\sqrt{N}$ with $N$ number of measurements; (iv) monitoring sensor temperatures; (v) enhance baffling to reduce ambient/stray light contamination; (vi) install a spectrograph with standard emission lines calibration lamp. Simultaneously, some questions regarding the stability of illumination need to be looked closer in the future. Repeatability of the throughput measurement must be tested by performing repeated scans at specific wavelengths and checking the reliability of the results. Systematic uncertainties can be kept below the statistical uncertainty threshold if concrete steps are taken toward the rigorous mitigation and assessment of each contribution mentioned in the paper. Additional investigation of the impact of beam spectral width on the filter's steep edges is also required.

Overall, the proposed system with detailed improvements has the potential to reach the systematic uncertainty limit of the CBP throughput curve measurement. Future work will focus on building the final refined system, adding the spectral quality evaluation in the analysis, and testing its "traveling" capabilities by calibrating the StarDICE experiment telescope at Observatoire de Haute-Provence. Having a calibration system that is transportable across different observatories is crucial for SNe Ia cosmology in the era of wide-field surveys collecting large amounts of data, where photometric calibration continues to be a significant limiting factor in enhancing cosmological constraints.

This paper is dedicated to the memory of our colleague Éric Nuss, who passed away in April 2023.

\section*{Acknowledgements}

This work received support from the Programme National Cosmology et Galaxies (PNCG) of CNRS/INSU with INP and IN2P3, co-funded by CEA and CNES and from the DIM ACAV program of the Île-de-France region.
Some of the results in this paper have been derived using the \textsc{scipy}, \textsc{numpy}, \textsc{lmfit}, and \textsc{astropy} packages. Figures in this article have been created using \textsc{matplotlib} and \textsc{corner}.
The authors thank the referees for their constructive comments that helped improve the quality and readability of the paper.

%%%%%%%%%%%%%%%%%%%%%%%%%%%%%%%%%%%%%%%%%%%%%%%%%%
\section*{Data Availability}
The data underlying this article will be shared on reasonable request to the corresponding author.
 
% The inclusion of a Data Availability Statement is a requirement for articles published in RASTI. Data Availability Statements provide a standardised format for readers to understand the availability of data underlying the research results described in the article. The statement may refer to original data generated in the course of the study or to third-party data analysed in the article. The statement should describe and provide means of access, where possible, by linking to the data or providing the required accession numbers for the relevant databases or DOIs.



\bibliographystyle{mnras}%rasti
\bibliography{bibliographie} % if your bibtex file is called example.bib

%\begin{thebibliography}{99}
%\end{thebibliography}

%%%%%%%%%%%%%%%%%%%%%%%%%%%%%%%%%%%%%%%%%%%%%%%%%%

%%%%%%%%%%%%%%%%% APPENDICES %%%%%%%%%%%%%%%%%%%%%

\appendix

\section{Beam spectrum}
\label{sec:appendix_spectra}
\begin{figure}
  \centering
  \includegraphics[width=1.0\hsize]{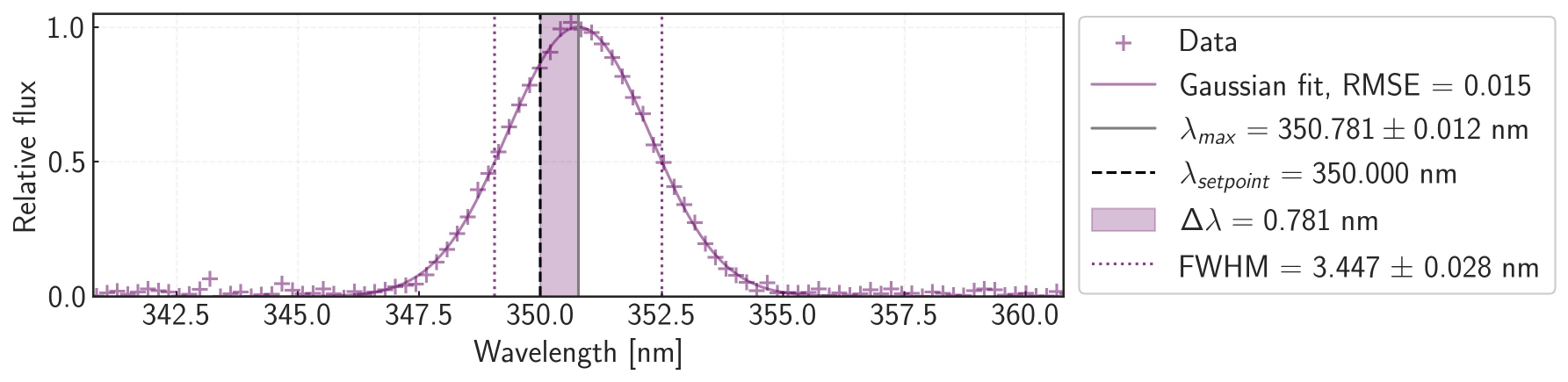}
  \includegraphics[width=1.0\hsize]{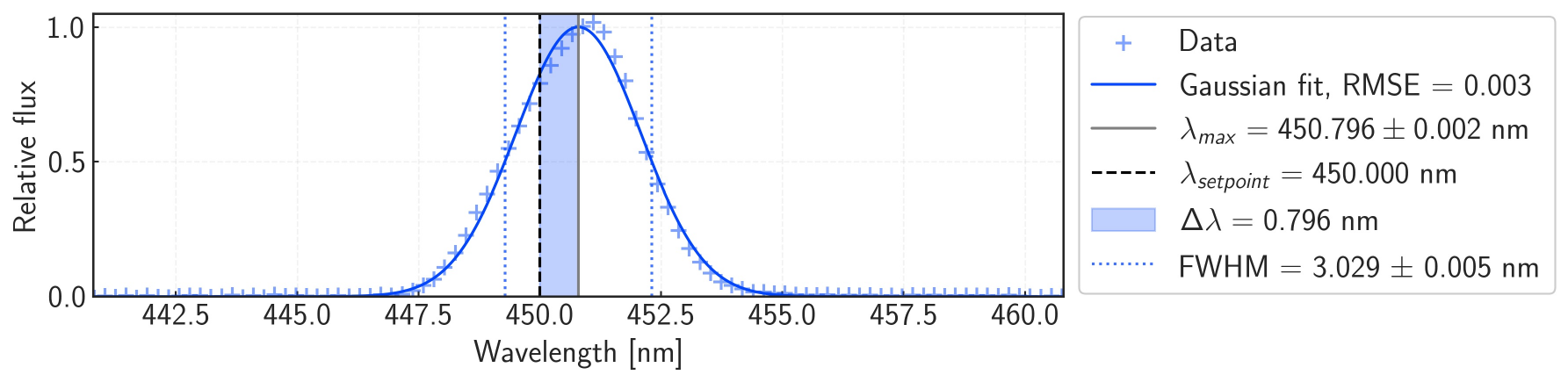}
  \includegraphics[width=1.0\hsize]{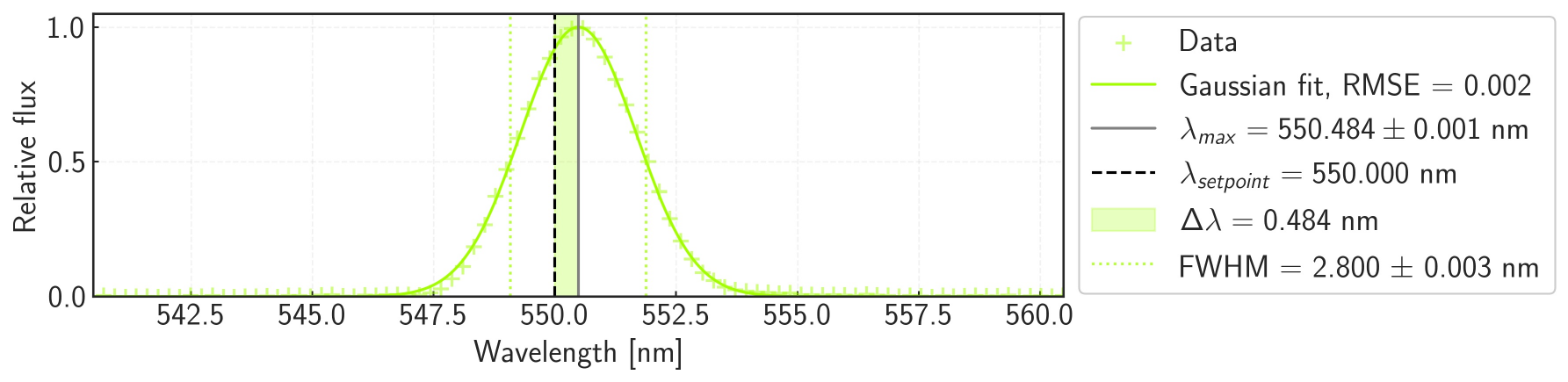}
  \includegraphics[width=1.0\hsize]{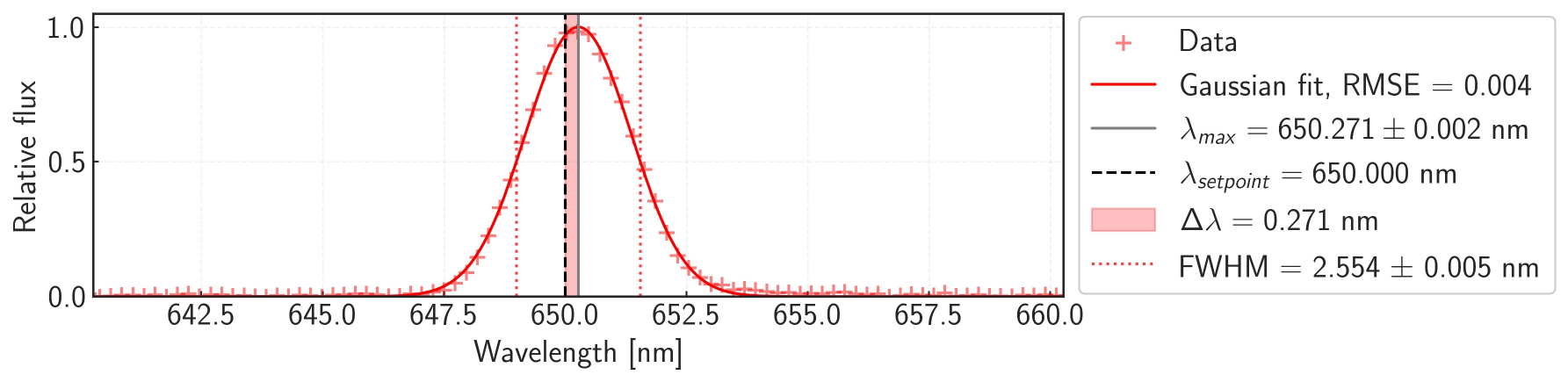}
  \includegraphics[width=1.0\hsize]{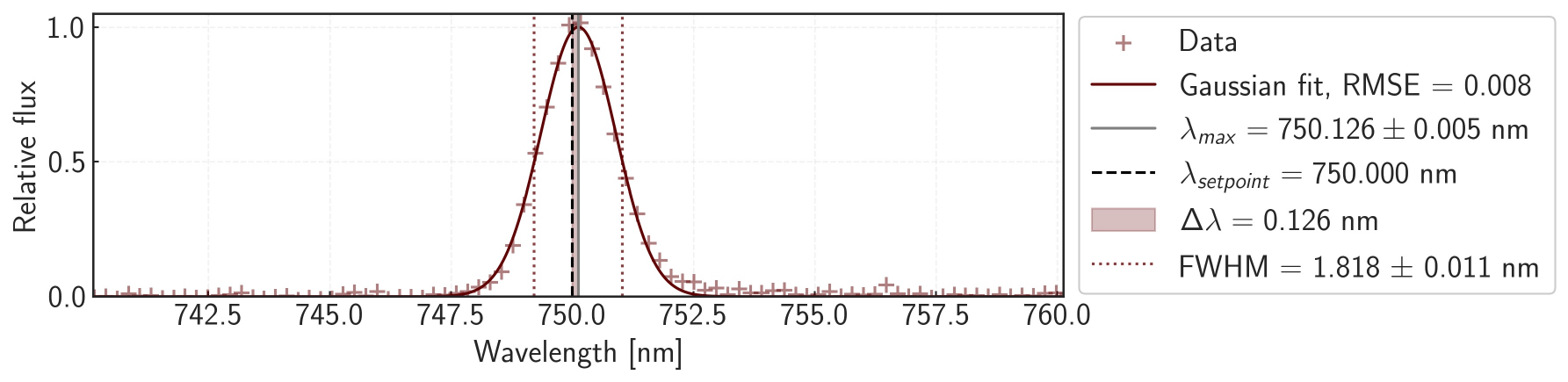}
  \caption{Spectrograph data with Gaussian fitting for 5 representative wavelengths. $\Delta\lambda$ shows the level of agreement between monochromator and spectrograph, while the FWHM gives the spectrograph's resolution.}
  \label{fig:spectra_350}
\end{figure}

For the proof of concept presented in this work, five spectra were taken with the spectrograph, at five different wavelengths spanning the range used in the analysis. As shown in Fig.~\ref{fig:spectra_350}, the line width decreases from 3.4 to 1.8 nm with increasing wavelength, and the monochromator/spectrograph agreement $\Delta\lambda$ is better than 0.8\,nm. Based on these points of measure, we corrected the monochromator wavelength based on the spectrograph spectra, using a 3rd order polynomial. Eventually, the spectrograph itself will be calibrated accurately and provide the nominal wavelength with excellent accuracy. As a result, we do not consider here any systematics related to wavelength calibration.
At this stage however, the spectrometer was not included in the automated script for data acquisition, and the spectrum acquisition was not systematic. Therefore, all wavelengths in this paper are not calibrated relatively to an absolute reference but only to the spectrograph reference to assess spectral beam central wavelength determination accuracy. Overall, we find excellent determination of the mean wavelength of the beam with accuracy staying below 0.015 nm and 0.8per cent relative uncertainty on the FWHM.

\section{Contours for simulations}

\begin{figure*}
  \centering
  \includegraphics[width=1.0\hsize]{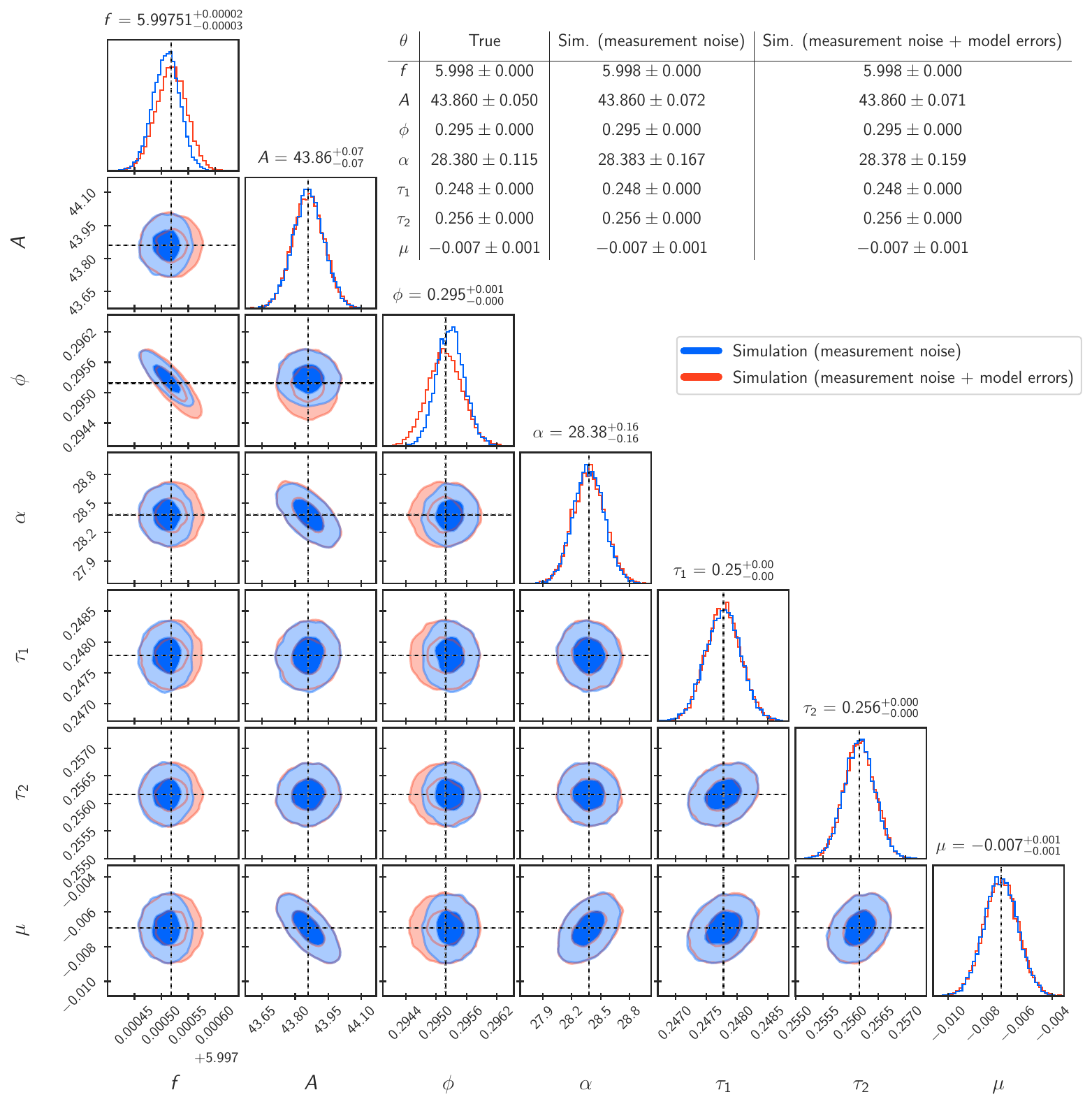}
  \caption{Posterior distributions for the fit of two mocks realizations of the PD simulated signal at $\lambda$ = 550 nm without (blue) and with (orange) residuals from the original observation and best-fitting curve. The colored contours show 1 and 2-$\sigma$ joint confidence intervals, the dotted black lines are the true values. The "True" column of the table presents the original best-fitting results on the observation.}
  \label{fig:contours_pd_sim_dual}
\end{figure*}

\begin{figure*}
  \centering
  \includegraphics[width=1.0\linewidth]{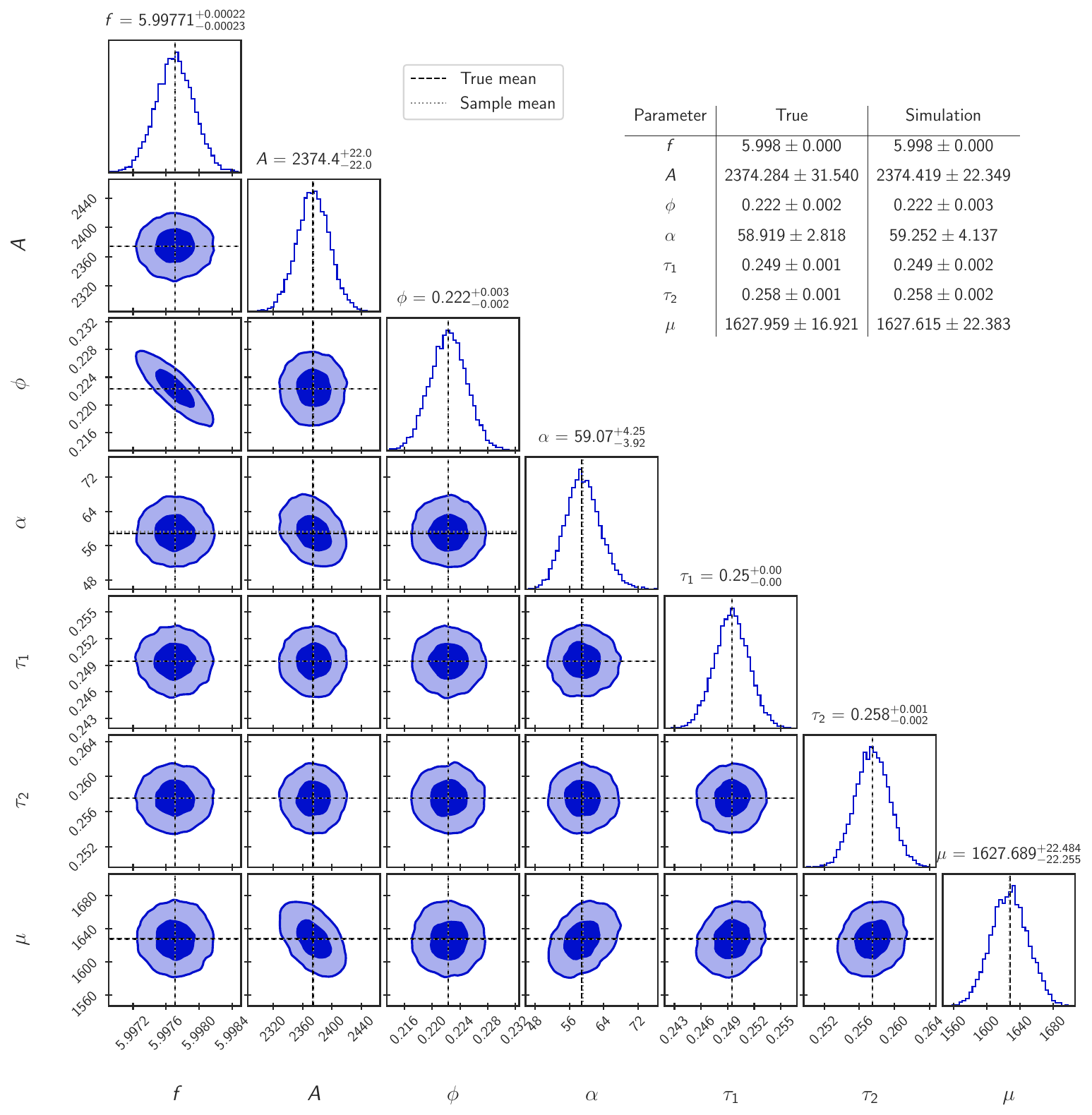}
  \caption{Posterior distributions for the fit of a single mock realization of the SC simulated signal at $\lambda$ = 550 nm. The blue contours show 1 and 2-$\sigma$ joint confidence intervals, the dotted black lines are the true values. The "True" column of the table presents the original best-fitting results on the observation.}
  \label{fig:contours_simulation_sc}
\end{figure*}

%%%%%%%%%%%%%%%%%%%%%%%%%%%%%%%%%%%%%%%%%%%%%%%%%%

% Don't change these lines
\bsp	% typesetting comment
\label{lastpage}
\end{document}